%% file: article.tex
\documentclass[]{interact}

\usepackage{epstopdf}
\usepackage{subfigure}

\usepackage{natbib}
\bibpunct[, ]{(}{)}{;}{a}{}{,}
\renewcommand\bibfont{\fontsize{10}{12}\selectfont}
\usepackage{amssymb}
\usepackage{amsmath}

\usepackage{graphicx}
\usepackage{svg}
\usepackage{epstopdf}

\usepackage{makecell}
\usepackage{multirow}
\usepackage{colortbl}
\usepackage{setspace} 

\newcolumntype{C}[1]{>{\centering\let\newline\\\arraybackslash\hspace{0pt}}m{#1}} 

\usepackage{placeins}

\usepackage{url}
\usepackage{hyperref}

\begin{document}

\articletype{Research Article}

\title{Surface solar radiation: AI satellite retrieval can outperform Heliosat and generalizes to other climate zones}

\author{
\name{K.~R. Schuurman\textsuperscript{a, b} and Angela Meyer\textsuperscript{a, b}\thanks{Corresponding author: angela.meyer@tudelft.nl}}
\affil{\textsuperscript{a}Geosciences and Remote Sensing, Delft University of Technology, Delft, The Netherlands; \textsuperscript{b}IDAS, Bern University of Applied Sciences, Biel, Switzerland}
}

\maketitle

\begin{abstract}

Accurate estimates of surface solar irradiance (SSI) are essential for solar resource assessments and solar energy forecasts in grid integration and building control applications. SSI estimates for spatially extended regions can be retrieved from geostationary satellites such as Meteosat. Traditional SSI satellite retrievals like Heliosat rely on physical radiative transfer modelling. We introduce a machine-learning-based satellite retrieval for instantaneous SSI and demonstrate its capability to provide accurate and generalizable SSI estimates across Europe. Our deep learning retrieval provides near real-time SSI estimates based on data-driven emulation of Heliosat and fine-tuning on pyranometer networks. 
By including SSI from ground stations, our SSI retrieval model can outperform Heliosat accuracy and generalize well to regions with other climates and surface albedos in cloudy conditions (clear-sky index $<$ 0.8). 
Our results indicate that the generalizability of a data-driven SSI retrieval model is not only related to the model training data or training method, but also depends on the amount of cloudiness present in the location at which SSI is retrieved with the data-driven model. 
We found that, in cloudy conditions, a model trained only on ground stations can estimate SSI accurately even in locations with different surface albedos, far away from the training test domain.
We also show that the SSI retrieved from Heliosat exhibits large biases in mountain regions, and that training and fine-tuning our retrieval models on SSI data from ground stations strongly reduces these biases, outperforming Heliosat. Furthermore, we quantify the relative importance of the Meteosat channels and other predictor variables like solar zenith angle for the accuracy of our deep learning SSI retrieval model in different cloud conditions. We find that in cloudy conditions multiple near-infrared and infrared channels enhance the performance. Our results can facilitate the development of more accurate satellite retrieval models of surface solar irradiance.

\end{abstract}


\section{Introduction}
Surface radiation estimates are required in solar resource assessments~\citep{yangWorldwideValidationEarth2022, carpentieriSatellitederivedSolarRadiation2023, vuilleumierAccuracySatellitederivedSolar2020}, solar forecast models~\citep{yangReviewSolarForecasting2022, carpentieriSatellitederivedSolarRadiation2023} and climate studies. 
High temporal and spatial coverage and resolution are essential for forecasting solar radiation for minutes to hours ahead (solar nowcasting) across large regions. While plant-level forecasts often rely on ground-based sensors~\citep{Nouri2019, KuhnSC2018, Barry2023}, forecasts across spatially extended regions require surface radiation retrieved from geostationary satellites
\citep{kosmopoulosMultiLayerCloudMotion2024, carpentieriExtendingIntradaySolar2025, knolDeepLearningSolar2021,  ayetNowcastingSolarIrradiance2018,kellerhalsCloudNowcastingStructurePreserving2022}.
Grid operators and energy providers use solar nowcasts to anticipate volatile solar energy and ensure grid stability. The relation between surface solar irradiance (SSI) and top-of-atmosphere reflectances measured from geostationary satellites has been explored and leveraged for retrieving SSI estimates for more than five decades \citep{tarpleyEstimatingIncidentSolar1979, mullerRemoteSensingSolar2022}. 
Satellite retrievals of SSI utilize top-of-atmosphere visible and infrared reflectances to estimate the amount of solar radiation that reaches the Earth's surface. 

Data-driven solar retrieval models can outperform physical models for single ground stations~\citep{cornejo-buenoMachineLearningRegressors2019}. Deep learning has enabled increases in SSI retrieval accuracy of up to 20\% compared to state-of-the-art retrievals~\citep{quesada-ruizAdvancedANNbasedMethod2015, haoEstimatingHourlyLand2019, huangEstimatingSurfaceSolar2019, verboisRetrievalSurfaceSolar2023}. 
Recent data-driven retrieval models have been trained on pyranometer ground stations with geostationarily measured radiances as predictors to retrieve SSI by either post-processing previous SSI retrievals~\citep{verboisImprovementSatellitederivedSurface2023, haoDSCOVREPICderivedGlobal2020, maEstimationSurfaceShortwave2020} or by directly estimating SSI retrievals for individual ground stations~\citep{verboisRetrievalSurfaceSolar2023, jiangDeepLearningAlgorithm2019, cornejo-buenoMachineLearningRegressors2019, quesada-ruizAdvancedANNbasedMethod2015, haoDSCOVREPICderivedGlobal2020, shiFirstEstimationHighresolution2023}.
SSI retrievals are expected to not only achieve high accuracy and low biases within the geographic region of their training set ground stations but also well beyond. In particular, an SSI retrieval should also be accurate in regions with other surface albedos and other climates, which have not been included in the development of the SSI retrieval model (out-of-domain generalization)~\citep{ haoDSCOVREPICderivedGlobal2020,verboisRetrievalSurfaceSolar2023, jiangDeepLearningAlgorithm2019, yangWorldwideValidationEarth2022}. Existing machine-learning based SSI retrievals were found to lack generalizability and to feature artefacts showing up in SSI estimates at locations outside their training set domains~\citep{verboisRetrievalSurfaceSolar2023,yangWorldwideValidationEarth2022}.

The goal of our study is to develop a machine-learning-based SSI retrieval model that enables accurate generalizable SSI estimates across all of Europe and North Africa at low retrieval latency time. 
We aim to go beyond previous work by providing an accurate data-driven satellite retrieval of instantaneous SSI that can generalize to unseen locations beyond the training set region. 
Our study provides the first machine learning satellite retrieval of instantaneous SSI, to our knowledge. Instantaneous SSI estimates are highly relevant for estimating and forecasting SSI at intra-hour and intra-day time scales.
We achieve improved generalizability by emulating a radiative-transfer-based SSI retrieval algorithm and fine-tuning it on ground station measurements.
SSI retrieval emulation involves the application of SSI fields, in our case from Heliosat, as target variables in the emulator training. Emulation is also commonly referred to as surrogate modelling.

The structure of this paper is outlined as follows. Section \ref{sec:previouswork} reviews existing SSI retrieval methods and concludes with the novelty of this paper. 
Section \ref{sec:data} introduces the datasets utilized in this study. The deep learning SSI retrieval and its development are detailed in section \ref{sec:retrievalmodel}. Results of this study are discussed in section \ref{sec:results} and conclusions are provided in section \ref{sec:conclusion}.

\section{Previous work}
\label{sec:previouswork}

One of the first SSI satellite retrieval algorithms, Heliosat-1, included an empirical formula to calculate SSI \citep{canoMethodDeterminationGlobal1986}. An empirical clear-sky model estimated SSI in the absence of clouds and assuming a given atmospheric composition. The transmissivity of the clouds was estimated by the ratio between top-of-atmosphere reflectivity measured and the minimum reflectivity over a month for cloud-free skies. The clear-sky reflectance mainly depends on the local solar zenith angle (SZA) and the surface albedo at the monitored site. Reliable surface albedo estimates are considered essential for accurate SSI retrievals~\citep{tournadreAlternativeCloudIndex2022}.
Current state-of-the-art surface solar irradiance retrieval algorithms, such as CAMS~\citep{schroedter-homscheidtSurfaceSolarIrradiation2022}, SARAH-3~\citep{pfeifrothSurfaceRadiationData2023}, MSG-CPP~\citep{roebelingCloudPropertyRetrievals2006, greuellRetrievalValidationGlobal2013}, and SolarGIS~\citep{suriSatellitebasedSolarResource2014}, use a combination of radiative transfer model (RTM) and the clear-sky index (CSI) to estimate SSI. 
The CSI is defined as the ratio between all-sky SSI and clear-sky SSI, $\text{CSI} = \frac{SSI_{all-sky}}{SSI_{clear-sky}}$, where the clear-sky SSI is calculated using a clear-sky RTM.
The latter performs corrections to the clear-sky SSI due to attenuation from aerosols and trace gases.
CAMS, SARAH-3 and MSG-CPP apply a cloudy-sky RTM to extract all-sky SSI based on cloud optical properties.
RTM simulations are computationally expensive, so all three methods utilize look-up tables created by calculating radiative properties for various cloud and aerosol conditions. 
The observed cloud optical properties, aerosol properties and surface albedo are interpolated within the conditions provided by the look-up table to estimate global surface solar radiation. 
In MSG-CPP, the cloud optical properties are derived from near-infrared, visible light and surface albedo as parameters to a cloudy RTM look-up table~\citep{roebelingCloudPropertyRetrievals2006, muellerNewAlgorithmSatelliteBased2012}. 
While the SSI retrieval algorithms differ, they are all based on the visible and partially near-infrared reflectivity from Meteosat Second Generation satellites. 

Recent research on deep learning (DL) for surface solar irradiance retrieval can be categorized into two main branches: post-processing of SSI derived from radiative transfer model and direct DL retrieval models.
In post-processing, DL is employed as a post-processing step for existing retrieval algorithms by using spectral imagery and derived variables as input to improve on the initial SSI retrieval estimate. 
\citet{cornejo-buenoMachineLearningRegressors2019} benchmarked a multitude of DL networks for post-processing Heliosat-2 and CAMS output, and found moderate improvements over a single station in Toledo, Spain. 
\citet{verboisImprovementSatellitederivedSurface2023} trained an extreme gradient boosting network to improve SSI estimates from HelioClim-3 \citep{blancHelioClimProjectSurface2011} based on SSI from a group of 283 pyranometers spread across France. 
\citet{maEstimationSurfaceShortwave2020} inferred instantaneous values of SSI (and other radiative transfer model outputs) by training a fully-connected neural network directly on an RTM with input level L2 products from the Himawari-8 satellite. 
They validated the model based on ground stations across Asia and Oceania and found an average instantaneous MBE and RMSE of 8.1 and 125.9 $Wm^{-2}$. 
These methods have not been validated with regard to their generalizability to other regions outside the training set domain.
Direct retrieval DL models, on the other hand, estimate SSI directly from visible or infrared satellite channels or ground station data. 
\citet{quesada-ruizAdvancedANNbasedMethod2015} trained three model ensembles of ten fully connected feed-forward neural networks each on all visible and infrared channels' pixel values of the Spinning Enhanced Visible and Infrared Imager (SEVIRI) plus a clear-sky SSI estimate and used SSI from eight ground stations in Europe as predictands (ground truth). 
Each ensemble was designed to estimate hourly SSI for different weather conditions: clear-sky (CSI $\geq$ 0.8), intermediate (0.8 $>$ CSI $\geq$ 0.4), and overcast conditions (CSI $<$ 0.4). A separate ensemble network of the same number of models predicted the clear-sky index. 
The method achieved an average RMSE skill score improvement of 17\% compared to the Heliosat-2 method \citep{rigollierMethodHeliosat2Deriving2004} across 28 validation stations.
In another study, \citet{jiangDeepLearningAlgorithm2019} applied a convolutional residual network to estimate hourly SSI based on the Multifunctional Transport Satellites (MTSAT) series with 16$\times$16 pixel patches of visible reflectances. 
The network was trained on 90 pyranometer stations spread out over different land surface classes and climates in China and inferred over the entirety of China. 
The network performance on eight validation pyranometers within the region showed an average RMSE of 93 $Wm^{-2}$. 
They did not compare those results with an existing radiative transfer-based retrieval algorithm.
\citet{jangEstimatingHourlySurface2022} created a CNN-based DL retrieval model trained on hourly averaged SSI measured at 81 weather stations from 2019 to 2020 across South Korea. They utilized an extra auxiliary variable the extraterrestrial solar radiation compared to \citep{jiangDeepLearningAlgorithm2019}. The reported MBE and RMSE for the same stations from 2020 to 2021 were -1.9 and 50.0 $Wm^{-2}$ compared to  32.8 and 90.1 $Wm^{-2}$ for the operational RTM model. 
Their validation on the same training stations does not allow any conclusions made for generalization over the entire domain.
\citet{shiFirstEstimationHighresolution2023} was the first to address the importance of different level-1 satellite inputs and auxiliary inputs to direct SSI retrieval using shapely values derived from the random forest model. 
They trained the model to predict hourly SSI values based on visible (0.55 - 0.77$\, \mu m$), near-infrared (3.5 - 4$\, \mu m$), infrared (6.9 - 7.3, 11.5 - 12.5$\, \mu m$) from the satellite FY-4A and auxiliary variables on ground stations across China.
The most important input was the solar zenith angle followed by the visible channel, solar azimuth angle, near-infrared and infrared (11.5 - 12.5$\, \mu m$) channel.
Compared to \citep{maEstimationSurfaceShortwave2020, jiangDeepLearningAlgorithm2019} the random forest model performed the worst with an hourly MBE and RMSE of -5.64 and 147.0 $Wm^{-2}$ over the ground stations of China.

\par
\citet{verboisRetrievalSurfaceSolar2023} were the first to explore the out-of-sample generalizability of DL models for SSI retrievals.  They trained a fully connected neural network on the hourly SSI from pyranometers over France incorporating 13 time steps of 15 minutes patches of 3$\times$3 from three SEVIRI channels: the visible $0.6\, \mu m$ and $0.8 \, \mu m$ (VIS006, VIS008) and infrared $10.8 \, \mu m$ (IR108).
They showed the performance of the DL network to be better than CAMS with an average RMSE skill score of 20\% within the training domain of southeastern France, but outside this limited domain, in the validation set over the rest of France, performance dropped off significantly to skill scores of -100\% to -40\% for some stations. 
The worse performing stations were located in coastal regions with a generally lower surface albedo which were not represented in the training dataset. 
\citet{verboisRetrievalSurfaceSolar2023} analyzed if average surface albedo impacted the generalization of the DL network and found a significant Spearman correlation of 0.346 (p $<$ 0.01) between surface albedo and RMSE skill score.
The results of \citet{verboisRetrievalSurfaceSolar2023} show challenges in generalising a DL retrieval to unseen locations far differentiated from the training set.
\par
This paper introduces an instantaneous direct retrieval model designed with two key objectives: ensuring fast inference and achieving robust generalizability across diverse and previously unseen climates and surfaces. 
We employ a convolutional neural network (CNN) architecture inspired by \citet{jiangDeepLearningAlgorithm2019}, utilizing only level-1 products from SEVIRI and auxiliary inputs that can be precomputed. We let it train on instantaneous SSI compared to hourly averaged SSI \citep{jiangDeepLearningAlgorithm2019, quesada-ruizAdvancedANNbasedMethod2015, shiFirstEstimationHighresolution2023, verboisImprovementSatellitederivedSurface2023}.
Initially, the model is trained to emulate a radiative transfer model (RTM) using the Heliosat-3 SARAH-3 dataset. 
The most similar methodology was implemented by \citet{maEstimationSurfaceShortwave2020} wherein they emulate an RTM instantaneous output but use level-2 products as inputs instead of directly retrieving the SSI from level-1.5 products here.
None of the previous works have first tried to emulate an RTM and subsequently fine-tuned the model using a ground station dataset that spans only a subset of the domain. 
This approach aims to address the generalizability challenges of deep learning-based retrieval methods by leveraging the broad applicability of the RTM while maintaining the high performance of a direct retrieval network trained on ground stations.

\section{Data}
\label{sec:data}

\subsection{Meteosat}
Visible and infrared radiances from Earth that are measured by geostationary satellites provide the basis of SSI retrieval products. The Meteosat Second Generation (MSG) satellite at $0^\circ$\,E, $0^\circ$\,N is monitoring the regions of interest in our study – Europe and North Africa – with the Spinning Enhanced Visible and Infrared Imager (SEVIRI) that scans from East to West by rotating its sensor's view. After each rotation, the sensor resets and scans the next line from North to South. 
One line scan is close to an instantaneous capture of longitudinal pixels, but individual scan lines have a delay associated with them, so an entire scan of the satellite's disk takes approximately 12 minutes~\citep{schmetzIntroductionMeteosatSecond2002}. 
The ground resolution of the sensor is 3$\times$3\,km at nadir.
All eleven visible and infrared channels are included as input to the emulator model training.

\subsection{Heliosat SARAH-3}
\label{sec:heliosat}
Heliosat SARAH-3 \citep{pfeifrothSurfaceRadiationData2023} provides Meteosat-derived estimates of surface solar irradiance and other atmospheric variables related to clouds and radiation. It spans Meteosat First Generation (1977-2017) and MSG (2002-present) satellite retrievals from their respective instruments MVIRI and SEVIRI. The MVIRI is the predecessor of SEVIRI.
SARAH-3 employs only the visible channels to keep the SARAH-3 record homogeneous in time.

The SARAH-3 algorithm draws on the cloud index method of the original Heliosat-1 algorithm by \citet{canoMethodDeterminationGlobal1986} to derive an estimated effective cloud albedo CAL. 
The cloud index method calculates the ratio between measured broadband reflectance $\rho$, clear-sky reflectance $\rho_{cls}$ and maximum reflectance $\rho_{max}$ as CAL $= \frac{\rho - \rho_{cls}}{\rho_{max} - \rho_{cls}}$. 
The clear-sky and maximum reflectance are estimated from the frequency distribution within a month, where the maximum reflectance is taken as the $95^{th}$ quantile in the month and the clear-sky reflectance as a stationary variate of lowest $\rho$ values. 
The lowest reflectance in the month is not taken as clear-sky reflectance due to cloud shadows which give low reflectance outliers in the tail of the distribution. 
The effective cloud albedo is converted to a cloud index with a smooth piecewise function defined in \citet{dagestadMeanBiasDeviation2004}. 
The observed pixel-wise cloud index serves as input to a  radiative transfer look-up table to retrieve SSI~\citep{muellerNewAlgorithmSatelliteBased2012}. 
Instantaneous SSI from SARAH-3 is published with a two-week delay at a frequency of 30 minutes, so every second SEVIRI capture.

\subsection{Grid data preprocessing}
The research domain of our study was selected as 29°\,N–62°\,N and 9°\,W–28°\,E to include most of Europe and parts of Northern Africa. Given the availability of high-quality measurements from IEA-PVPS in the years 2016-2022, we chose SARAH-3 and SEVIRI images from 2016 to 2023 for this study. 
The SEVIRI Level-1.5 data was acquired from the EUMETSAT datastore. We utilized eleven visible and infrared SEVIRI channels. The SEVIRI Level-1.5 data corresponds to image data that is corrected for unwanted radiometric and geometric effects. 
The Level-1.5 VIS006, VIS008, and IR016 reflectivity radiances were converted to bidirectional reflectance factor.  The Level-1.5 emissivity radiances for channels IR039, WV062, WV073, IR087, IR097, IR108, IR120 and IR134 were converted to brightness temperature using the Satpy module~\citep{martin_raspaud_2024_12925793}. 
The conversions homogenized the record between different MSG satellites and corrected the reflectivity channels for changing Sun-Earth distance.
The images were reprojected to the geographic SARAH-3 grid with the nearest neighborhood interpolation. 

\subsection{Ground stations}

Our SSI emulator models were validated on ground-based SSI sensors. Pyranometers measure the total amount of solar radiation from a half-dome in segments of one to ten minutes. 
These sparse ground observations represent only a single point in space whereas the SEVIRI imager measures an average reflectivity over a ground area of approximately 3$\times$3\,km. 
The ground station datasets for validating the SSI emulators are summarised in Table \ref{tab:groundstations}. They include SSI measurements from the Royal Netherlands Meteorological Institute (KNMI), the German Weather Service (DWD) and the Swiss Federal Office for Meteorology and Climatology (MeteoSwiss).
Additionally, SSI measurements from the International Energy Agency Photovoltaic Power Systems Programme (IEA-PVPS) worldwide benchmark were gathered. They contain quality controlled SSI measurements from across Europe and North Africa~\citep{forstingerWorldwideSolarRadiation2022}. 
The IEA-PVPS dataset comprises 14 stations from different operators.
Six stations – CAB, CAR, CEN, PAL, PAY, and TOR – form part of the Baseline Surface Radiation Network (BSRN, \citep{essd-10-1491-2018}). 
Two stations – TAB and MIL – belong to the Centre for Energy, Environmental and Technological Research (CIEMAT) and Research on Energy Systems (RSE), respectively. 
The Swedish Meteorological and Hydrological Institute (SMHI) provided two additional stations, NOR and VIS. 
Finally, the EnerMENA project contributed four stations located in Northern Africa: GHA, MIS, OUJ, and TAT \citep{schulerEnerMENAMeteorologicalNetwork2016}.
All stations of the IEA-PVPS dataset are equipped with three thermophile radiometers measuring SSI, direct normal irradiance and diffuse irradiance at 1-minute intervals~\citep{forstingerWorldwideSolarRadiation2022}.  

\begin{table}[ht!]
    \tbl{Datasets of SSI ground station measurements. All years are included in the training/validation.}{
    \begin{tabular}{lcccc} \toprule
        & \multicolumn{4}{c}{Details} \\ \cmidrule{2-5}
         Dataset&  Quality control & Stations& \makecell{Measurement \\ frequency (minutes)}& Years\\ \hline  
         IEA-PVPS& 
     Yes & 14 & 1 & 2016-2022\\   
    DWD &  Automatic & 99 & 10 & 2016-2022\\ 
    KNMI & No & 34 & 10 & 2016-2022\\ 
    MeteoSwiss& Automatic & 135 & 10 & 2016-2022\\ \bottomrule
    \end{tabular}
    }
    \label{tab:groundstations}
\end{table}

\begin{figure}[h!]
    \centering
    \includegraphics[width=.7\textwidth]{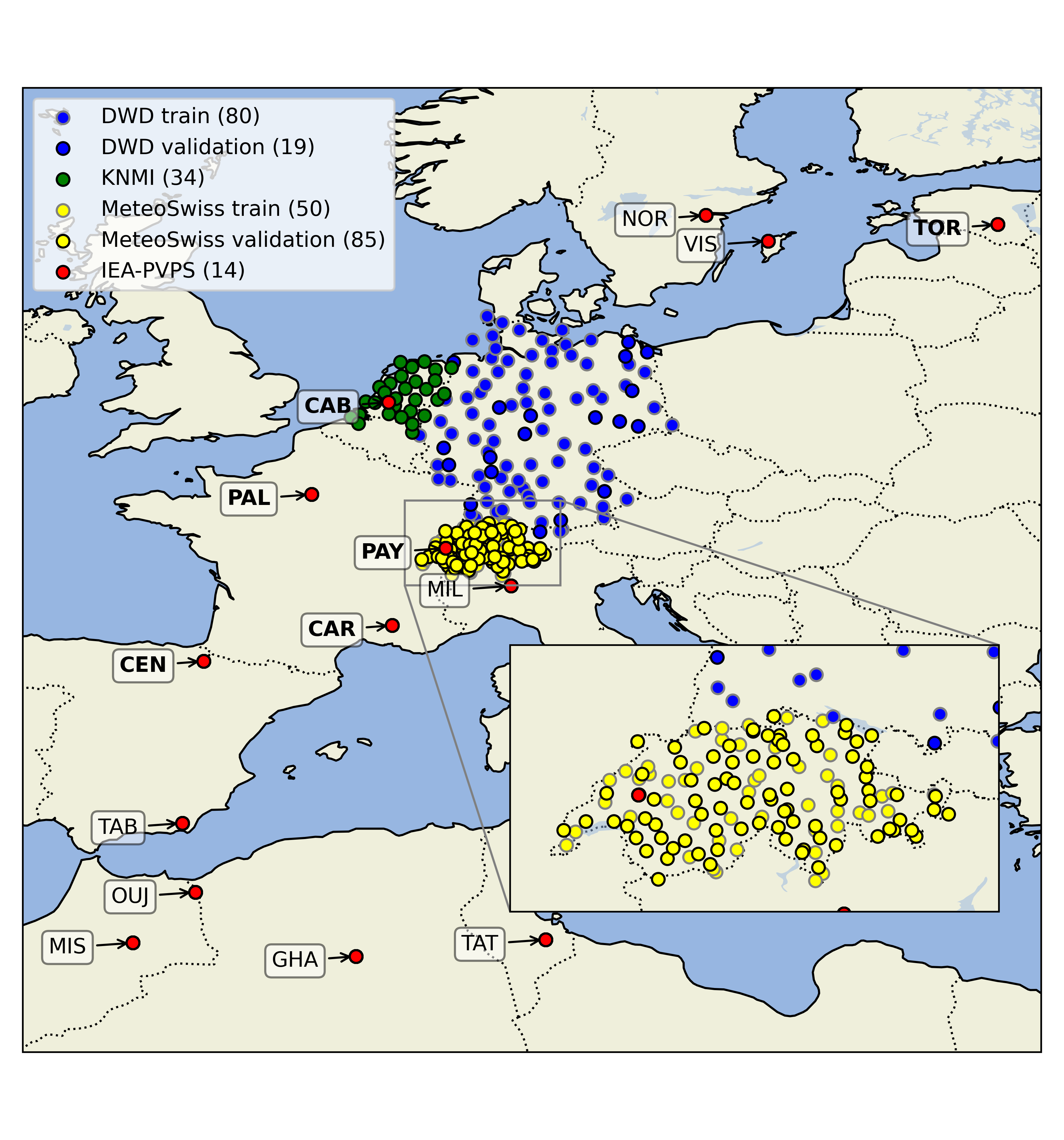}
    \caption{Ground station measurements collected for this study. Bold-named stations are part of the BSRN network. Gray-bordered stations indicate sites within the DWD and MeteoSwiss training datasets. The domain shown corresponds to the domain of this study, 29°N–62°N and 9°W–28°E.}
    \label{fig:pyranometer_locations}
\end{figure}

The DWD dataset comprises SSI measurements from 99 radiometers including 70 scanning pyrheliometers (ScaPP) and 19 Kipp \& Zonen CM11 or CM21 pyranometers~\citep{beckerQualityAssessmentHeterogeneous2012}. The ScaPP instruments operate within a wavelength range of 0.3 to 1.1$\,\mu m$ based on a silicon detector. 
The MeteoSwiss dataset includes SSI measurements from 135 weather stations across Switzerland, all equipped with Kipp \& Zonen CM21 pyranometers, with multiple stations located in Alpine regions.
The KNMI dataset consists of 34 automatic weather stations, each featuring a single radiometer. This dataset is provided without quality control.
All measurement stations are depicted in Figure \ref{fig:pyranometer_locations}.

Due to its extensive spatial coverage and stringent quality control, the IEA-PVPS dataset served as the reference dataset for validating models and comparing their performance. The other datasets receive less weight in the analysis. The exact date ranges of observations may vary across ground stations to some extent for each of the four datasets (Table \ref{tab:groundstations}) because some stations were commissioned or decommissioned during the timeframe. Filtering on quality control can leave significant gaps in the data record as well.

\section{SSI retrieval model}
\label{sec:retrievalmodel}

\begin{figure}[]
    \centering
 \includegraphics[width=1.0\textwidth]{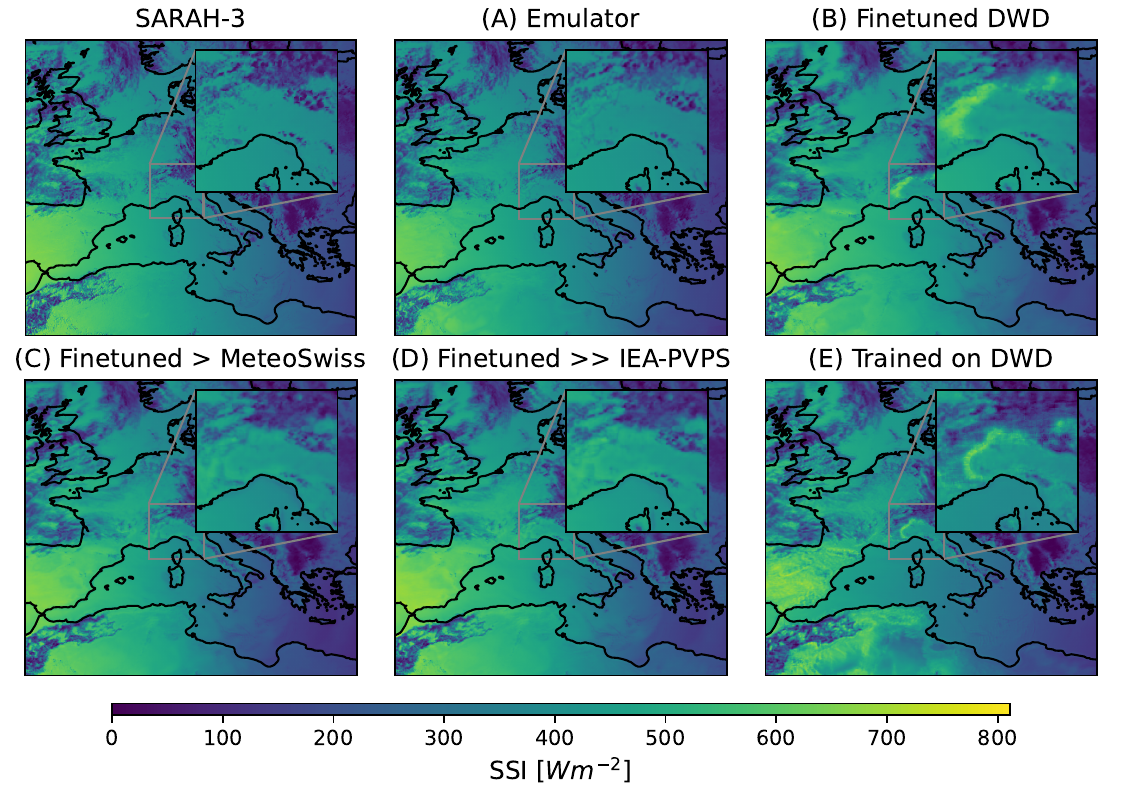}
    \caption{Model comparison for the instantaneous SSI retrieval on 2022-08-20 at 15:30 UTC.}
    \label{fig:image_comparison}
\end{figure}

We trained five retrieval models in total. All models were trained on the same predictors: The Meteosat visible and infrared channels and the above scalar variables.
The models are: E) a model trained only on DWD ground station data as target variable (section \ref{sec:baselinetraining}); A) an emulator model trained on the Heliosat SARAH-3 SSI as surrogate (section \ref{sec:emulatortraining}); and B-D) Three fine-tuned models, which trained to emulate SARAH-3 first and then fine-tuned on ground station data (section \ref{sec:finetuningtraining}). 
The models are summarized in Table \ref{tab:models}. An example case of SSI retrieved across the entire domain is shown in Figure \ref{fig:image_comparison} for Heliosat and all five models.

The following first section describes the network architecture for all models. 
Section \ref{sec:baselinetraining} details the training on ground station measurements and introduces the SSI retrieval model. Section \ref{sec:emulatortraining} details the emulator training with SARAH-3 as a target variable and lastly section \ref{sec:finetuningtraining} shows the subsequent fine-tuning of the emulator on ground station data.

\subsection{Architecture}

We implemented a convolutional residual network (ConvResNet) similar to the one of \citet{jiangDeepLearningAlgorithm2019} with the following adjustments, as shown in Figure \ref{fig:convresnet} in the Appendix. 
Instead of 16$\times$16 we defined patches of 15$\times$15 pixels so that the centre pixel corresponds to the retrieved SSI pixel using all 11 visible and infrared channels. 
The output of our network is instantaneous SSI instead of hourly averaged SSI.
\citet{jiangDeepLearningAlgorithm2019} argued against including the solar zenith and azimuth angles in the network, suggesting that the spatial information is inherently captured by the convolutions. However, we find no compelling reason to exclude these variables, as they are strongly correlated with SSI and can be easily precomputed.
What's more, is that we applied batch normalization between each convolutional layer to prevent overfitting of the model.
Lastly, extra geographical information was given by the digital elevation model (DEM) from Copernicus GLO-90~\citep{airbusCopernicusDigitalElevation2024} coarsened to the same grid as MSG.

The first part of the CNN is the feature mapping part where the spatial size is kept constant within the first two convolutional mappings (Figure \ref{fig:convresnet}).  
Channel features were extracted from the patches by the CNN. 
Further features included the day of the year, the latitude, longitude, solar zenith angle and solar azimuth angle and were appended to the channel features and fed through a two-layer fully-connected neural network (FCN). 
The second part consists of a 2$\times$2 max pooling (MP) step to reduce spatial size followed by a convolutional residual block. 
The third part is the same where a 2$\times$2 MP is followed by a convolutional residual block. The output of the CNN is first global average pooled (GAP) and then flattened to serve as input to the FCN. 
The kernel size in each convolutional layer is 3$\times$3. Each convolutional layer in the network is followed by a 2D batch normalisation and ReLU activation function. 
Each of the two FCN layers is followed by a 1D batch normalisation and ReLU activation. 

The scalar output for the DL network is the SSI and was clipped for validation and inference to zero when estimating negative SSI.
The input features SZA and solar azimuth angle (AZI) were computed at DEM altitude and nominal capture time. 
The retrieval with ConvResNet takes 15 seconds to run on a single Nvidia Tesla P100 GPU for a domain size of Europe (658$\times$736 pixels). The retrieval could be linearly parallelized for larger domains or lower inference times. A newer and faster GPU than the Nvidia Tesla P100 (2016) would speed up the inference as well.

\subsection{Training on ground stations}
\label{sec:baselinetraining}

Colocated samples in time and space were created between the ground station and images to train a retrieval model on ground stations.
The IEA-PVPS ground observations were averaged to 10 minutes from the start of each nominal satellite capture time. 
For the ground observations from KNMI, DWD and MeteoSwiss which already measured 10-minute averages, the SEVIRI nominal capture times were matched with ground observation measurement start times. 
This meant that non-quarterly observation times at the 10/20/40/50 minute marks were disregarded.
All available matching observations between 2016 and 2022 were used. 
Patches of 15$\times$15 pixels with the center pixel overlapping the ground station were selected.
Scalar input features latitude, longitude, SZA and AZI were calculated for the exact locations of the respective stations. 
The DWD stations were split randomly into 80 training and 19 validation stations. 
The density of the MeteoSwiss network is so high that we could afford to keep more validation stations than training stations, and they were randomly split into 50 training and 85 validation stations.
The KNMI stations were left out of training due to the lack of quality control.
The station data were filtered for SZA $<$ $85^\circ$ to select daytime periods only. 
When training on ground stations, the samples of different sites were first merged and then randomly shuffled to ensure that training, validation and test sets have similar statistical characteristics.
The model (E) was trained only on the DWD dataset for 15 epochs with a mean squared error loss with an Adam optimizer with weight decay \citep{kingmaAdamMethodStochastic2014}.

\subsection{Emulator training on Heliosat}
\label{sec:emulatortraining}
The emulator model (A) was trained on Heliosat SARAH-3 SSI. SARAH-3 provides SSI fields for every second capture of the SEVIRI imager.
The SARAH-3 SSI as target variable and the SEVIRI images and scalar features (latitude, longitude, SZA, AZI, day of the year) as predictor variables were merged at nominal capture time to form a dataset to train, validate and test our SSI emulator model.  The scalar features were calculated at the center of each pixel.
We split the SEVIRI images and SARAH-3 retrievals into training (years 2016--2021), validation (2022) and test set (2023).
Daytime locations were defined as pixels where SZA $< 85^\circ$.
Images with less than 50\% daytime pixels were dropped. 
The images were shuffled randomly for training.
A sample consists of 15$\times$15 patches of each grid input and the five scalar features from the center pixel.
The inputs and outputs were min-max normalized for all training and fine-tuning steps. 
A training batch consisted of 2048 uniformly distributed samples per image.
Only the daytime pixels were sampled from images containing both day- and nighttime pixels.  
A single epoch consisted of cycling through all images with 2048 samples per image. The network was trained for two epochs on a mean squared error loss with an Adam optimizer with weight decay \citep{kingmaAdamMethodStochastic2014}. 
The validation loss was monitored to prevent overfitting the training dataset. 
A learning rate of $10^{-4}$ dropping to $10^{-5}$ in the second epoch yielded a balance between training time and  recognizing of overfitting.
Every 10\% of a full training epoch, the equivalent of half a year of SEVIRI captures, the validation metrics were calculated and a checkpoint was made. From those checkpoints, the best model was picked.

\subsection{Fine-tuning on ground stations}
\label{sec:finetuningtraining}

The pre-trained emulator (A) was further fine-tuned on ground station data, yielding our three fine-tuned retrieval models (B-D). The fitted weights found in section \ref{sec:emulatortraining}, optimized to emulate Heliosat SARAH-3 SSI, were the starting point for fine-tuning the first retrieval model (B).
Fine-tuning in machine learning is the process of adapting a pre-trained model to statistically similar but somewhat different target datasets and tasks. While, in principle, that somewhat different dataset could be directly trained on, training a large model from scratch on a comparatively small dataset, such as a ground station dataset, risks overfitting: the model might learn to perform well on the training examples but generalize poorly to new data and locations. 
In the context of SSI retrieval, 
the ground stations are available over a limited subregion while the Heliosat SARAH-3 SSI is available across the entire SEVIRI disk. Here, the goal of fine-tuning is to increase the SSI retrieval's precision to ground station measurements without becoming biased toward them.

We applied three techniques to prevent the fine-tuned emulator from overfitting the ground station observations.
Firstly, the validation loss and bias on the original Heliosat SARAH-3 data were monitored every tenth of a training epoch. 
The frequency was adjusted depending on the number of samples in the fine-tuning set.
Secondly, a proximal loss function for the weights was added, $ \mathbf{L}_{prox}(\theta, \theta_0) = \frac{\alpha}{2} || \theta - \theta_0 ||_2 $, 
which punishes the model for deviating from the weights learned in the previous domain. 
The $\theta_0$ weights specify the best-fit model for the Heliosat SARAH-3 emulation task and $\theta$ the weights after fine-tuning the task of estimating SSI for ground stations. 
The proximal loss intends to regularize overfitting on the ground observation constrained by the best-fit model for the emulation task (model B). 
Thirdly, the FCN part of the model was frozen, meaning the FCN weights were fixed for fine-tuning. This aims to keep the relationship between SZA and SSI stable because the SZA is only appended to the model inputs before the FCN.

In total, we trained three fine-tuned models (rows B, C and D in Table \ref{tab:models}).
To arrive at the first fine-tuned retrieval model (B), we fine-tuned the emulator on the DWD ground stations training set with a proximity loss ($\alpha = 5000$), a frozen FCN and a decreased learning rate $lr=10^{-6}$.
In the second stage of fine-tuning, which yielded the second fine-tuned model (C), the previous model fine-tuned on DWD (B) was further fine-tuned on the MeteoSwiss training set with a higher proximity loss ($\alpha = 10000$), a frozen FCN and the same learning rate.
In the last stage of fine-tuning, which yielded the third fine-tuned model (D), the model previously fine-tuned on DWD and MeteoSwiss ground stations (C) was further fine-tuned on the IEA-PVPS SSI measurements without a frozen FCN and with a smaller learning rate lr$=10^{-7}$.
For each stage, the previous model weights were set as the $\theta_0$ in the proximity loss, as shown in column 'Weights $\theta_0$' in Table \ref{tab:models}.
In the first two stages of fine-tuning, the validation loss and MBE per station on IEA-PVPS were monitored ten times per epoch to signalize a change in bias or overfitting to the stations on which the model was being fine-tuned. In the third stage, the validation loss and MBE on MeteoSwiss and DWD validation stations were monitored for the same reason. 
Each fine-tuning stage lasted at most two epochs on their respective training set or shorter when the model showed signs of overfitting or bias shifts.

\begin{table}[ht]
\tbl{Specification of the different trained models.}{

\input{models_table}}
\label{tab:models}
\end{table}

\FloatBarrier

\section{Results and discussion}
\label{sec:results}

\subsection{Characterisation of the SSI emulator}

\begin{figure}[h!]
    \centering  \includegraphics[width=.5\textwidth]{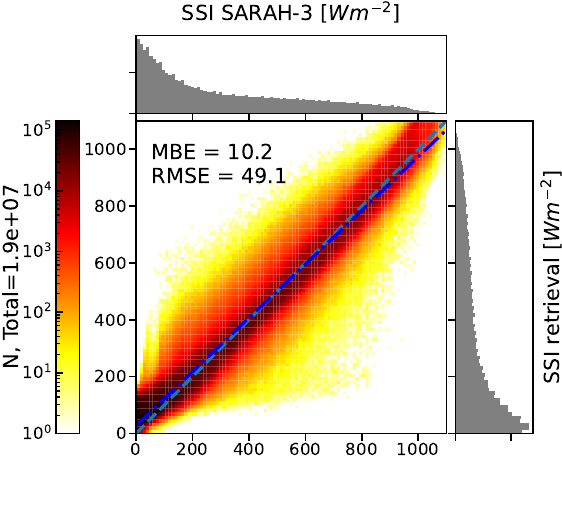}
    \caption{Scatter plot of the instantaneous SSI ('SSI retrieval') compared to the SSI of the Heliosat SARAH-3 validation set ('SSI'). The dashed lines indicate the 45° line (green) and a linear regression fit (blue), respectively.  The plot comprises 19.3 million SSI retrievals across Europe and North Africa from 2022.}
\label{fig:QQplot_validation_SARAH}
\end{figure}

\begin{figure}[t!]
\centering
    \subfigure[]{%
        \resizebox*{7cm}{!}{\includegraphics{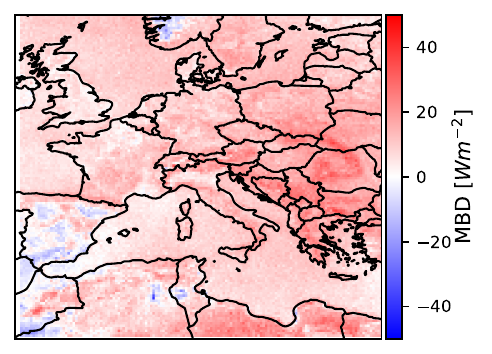}}}
    \subfigure[]{%
        \resizebox*{7cm}{!}{\includegraphics{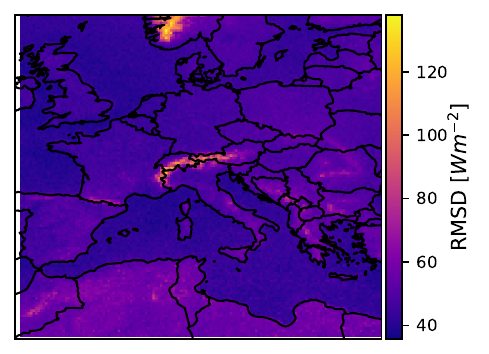}}}
\caption{Instantaneous mean bias deviation (a) and instantaneous root-mean-square deviation (b) of instantaneous SSI estimates of the emulator model with regard to SARAH-3 SSI averaged pixel-wise during daytime periods of 2023 on the test set. Samples were binned in  $0.25^\circ$x$0.25^\circ$ pixels to estimate regional MBDs and RMSDs. }
\label{fig:regional_analysis}
\end{figure}

The emulator model trained in section \ref{sec:baselinetraining} was compared to the Heliosat SARAH-3 SSI on the test set and to the ground stations SSI to characterise its performance. Instantaneous root-mean-square deviation (RMSD) and mean bias deviation (MBD) were only calculated pixel-wise and over daytime periods.
We found that the emulator follows the SARAH-3 SSI remarkably closely, as shown in Figures \ref{fig:QQplot_validation_SARAH} and \ref{fig:regional_analysis}. It emulates SSI with high accuracy and with biases negligibly small for solar energy applications such as resource assessments and forecasting. Specifically, the emulator model has a small positive MBD of 10.2\,$Wm^{-2}$ by somewhat overestimating low Heliosat SARAH-3 SSI values (Figure \ref{fig:QQplot_validation_SARAH}). The emulator generates an RMSD of 51.7\,$Wm^{-2}$ with regard to the SARAH-3 SSI on the test set.

The RMSD with regard to SARAH-3 is the largest in mountainous regions such as the Alps, the Pyrenees, and the Scandinavian mountains, with RMSD values of up to 125\,$Wm^{-2}$. 
Stations in mountainous regions tend to be affected by larger SSI retrieval uncertainties due to complex terrain, local clouds and snow cover~\citep{carpentieriSatellitederivedSolarRadiation2023}. 
Snow cover can be challenging to distinguish from clouds due to the similarity of visible reflectivity, which leads to underestimation when snow cover is mistaken for clouds~\citep{carpentieriSatellitederivedSolarRadiation2023}.
Non-moving highly reflective pixels are flagged and treated either as fog or snow cover by Heliosat SARAH-3~\citep{pfeifrothSurfaceRadiationData2023}. 
This inadvertently can lead to the opposite problem of overestimating SSI where it accurately identifies snow cover but does not consider clouds covering a snowy region. 

Deviations with regard to Heliosat SSI are also somewhat elevated across North Africa with RMSD values of around 75 $Wm^{-2}$.
The lowest regional RMSD of around 45 $Wm^{-2}$ can be found over large water bodies, such as the North Sea, the Atlantic and the Mediterranean Sea.
The common denominator for elevated RMSD is the persistent surface albedo during the year. Snow cover but also the bare surfaces of North African Sahara show increased reflectivity most of the year~\citep{bierwirthSpectralSurfaceAlbedo2009}. Water bodies on the contrary show a low amount of backscatter for most incidence angles and thus have generally low albedo~\citep{jinNewParameterizationSpectral2011}. 
A high uncertainty of the Heliosat method comes from large clear-sky reflectances~\citep{pfeifrothSurfaceRadiationData2023}. 
Large clear-sky reflectances lower the contrast between clear-sky and cloudy reflectance which linearly impacts the uncertainty in CSI and subsequently SSI of SARAH-3.
The heightened RMSD in these regions might be mostly due to the ground truth being noisy or inaccurate.

\subsection{Retrieval over Europe}
\label{sec:imagecomparison}

Figure \ref{fig:image_comparison} illustrates differences and similarities across our five SSI retrieval models (Table \ref{tab:models}) and Heliosat SARAH-3 on a randomly selected day and time.
Overall, the five retrieval models mimic all cloud features seen in SARAH-3 at remarkable quality. 
When examining the retrievals closely, some differences become apparent.
Firstly, the boundaries between clouds and clear-sky SSI tend to be somewhat sharper in SARAH-3 whereas the deep learning models exhibit moderate smoothing compared to SARAH-3. This may be attributed to the 16$\times$16 pixel inputs which likely blend the SSI with neighboring pixel information. However, it is important to note that determining which level of smoothness is more accurate is challenging due to the absence of spatially extensive SSI measurements over large areas as ground truth.


Secondly, both SARAH-3 and the emulator reveal a gradient from high SSI in the west to low SSI in the east in clear-sky regions. This gradient, which is caused by varying incident angles of sunlight, appears more gradual in SARAH-3 compared to our five data-driven models. For example, the emulator shows moderate deviations in this gradient over the Iberian Peninsula and Northern Africa, while models A and C-E display more pronounced deviations in the clear-sky gradient.
We also note that the emulator-based models (A--D) appear to exibit fewer artefacts than the model trained on only ground stations (E) which features spatially correlated structures of excessive SSI in the Alps, across the Iberian Peninsula and on the western parts of North Africa. These artefact structures appear to be correlated with orography. 

\subsection{Validating SSI retrieval models on ground stations}
\label{sec:validation}
\begin{figure}
    \centering
    \makebox[\textwidth][c]{
    \includegraphics[width=1\textwidth]{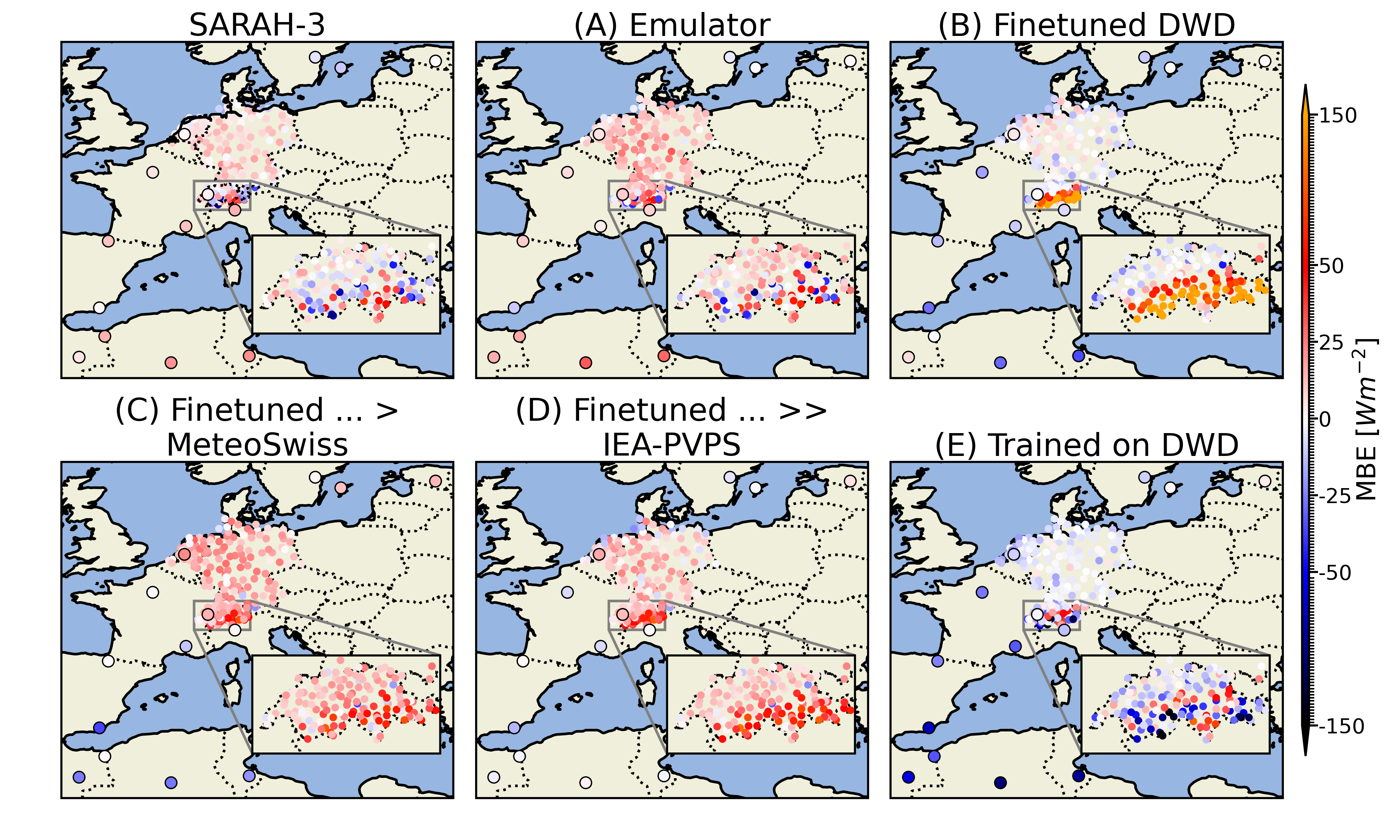}}
    \caption{The instantaneous Mean Bias Error (MBE) on all observation ground station sets, averaged over all observations with a SZA $< 85^\circ$.}
    \label{fig:scatter_mbe}
\end{figure}

\begin{figure}
    \centering
    \makebox[\textwidth][c]{
    \includegraphics[width=1\textwidth]{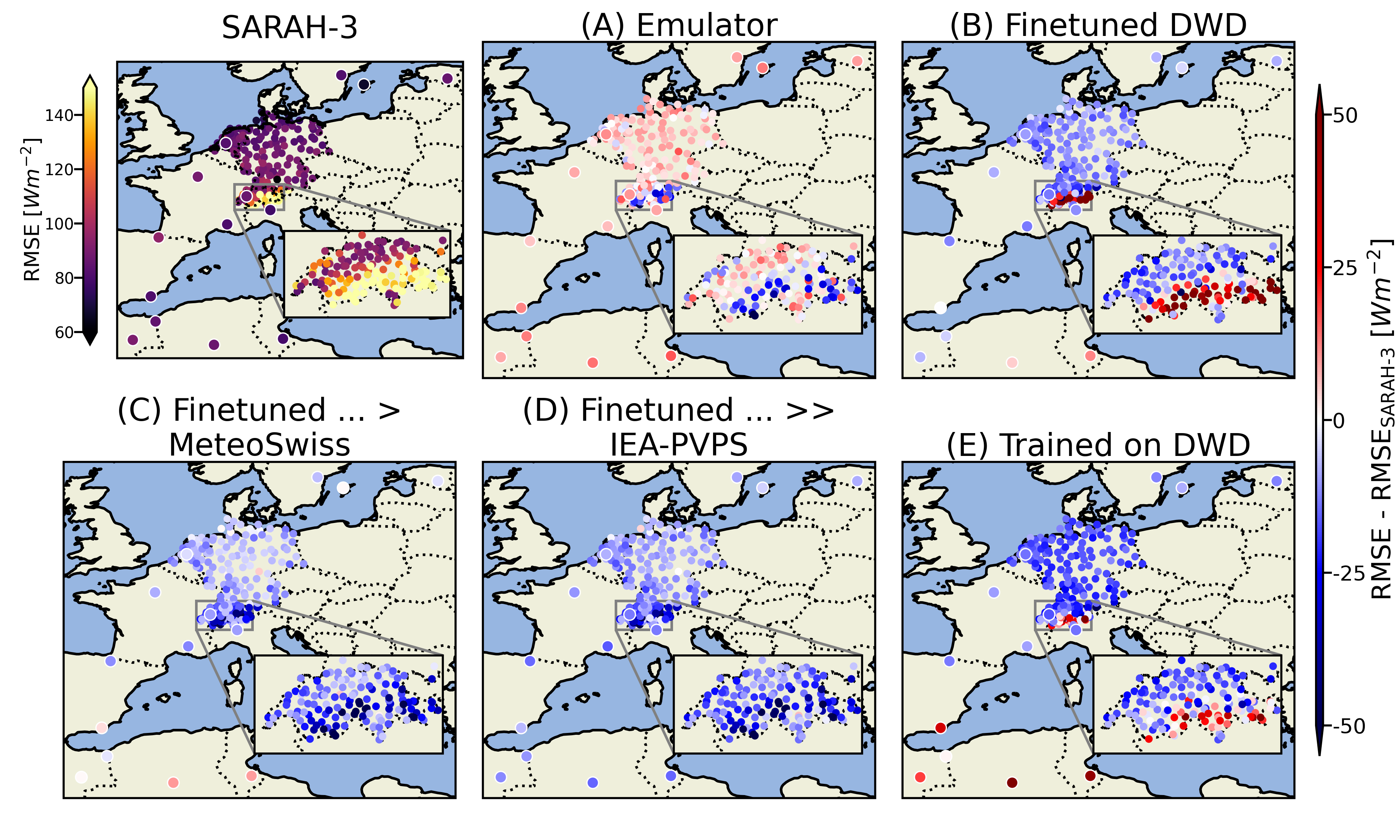}}
    \caption{The instantaneous Root-Mean-Square Error difference (RMSE$_{\text{diff}}$) between the SSI estimated by our retrieval models and Heliosat SARAH-3, averaged over all observations per ground station with a SZA $< 85^\circ$.}
    \label{fig:scatter_rmse}
\end{figure}

We validated our SSI retrieval models on multiple ground station networks across Europe and North Africa. To compare the models to actual SSI measurements rather than Heliosat SSI, we investigated root-mean-square errors (RMSE) and mean bias errors (MBE) with regard to the ground stations. 
The RMSE and MBE were calculated as averages over all matched ground station measurements from 2016 to 2022 including training set. 
Nominal start times HH:15 UTC and HH:45 UTC of the SEVIRI scan were excluded from the RMSE and MBE analysis to match the 30-minute frequency of the Heliosat SARAH-3 record.
In Figures \ref{fig:scatter_mbe} and \ref{fig:scatter_rmse}, the MBE and the difference in RMSE with regard to SARAH-3 were plotted by station.
The IEA-PVPS dataset provides the largest coverage of the considered domain and is used for the analysis. 
The numeric values of MBE and RMSE for the IEA-PVPS dataset are given in Tables \ref{tab:RMSE} and \ref{tab:MBE}. 

As shown in Figure \ref{fig:scatter_mbe}, the MBE of the SARAH-3 SSI is on the order of -10 to 10$\, Wm^{-2}$ except for mountainous sites. The emulator generates somewhat higher biases than SARAH-3. 
Fine-tuning on DWD stations accomplished a bias reduction compared to SARAH-3 SSI on a majority of DWD stations but also increased the biases in mountain regions. 
Subsequently fine-tuning the emulator fine-tuned on DWD also on the MeteoSwiss stations reduced the biases in the Alps, but left the biases across the southern sites at a similar level.
The accuracy and generalization of the fine-tuned model can be further increased by fine-tuning it also on the IEA-PVPS stations, yielding model E). This reduces the biases in the North African sites to near zero without affecting the biases across the KNMI, DWD and MeteoSwiss validation sets. However, note that due to the small number of ground stations in the IEA-PVPS dataset, fine-tuned model E was fine-tuned and validated on the same IEA-PVPS stations, so model E) may to some extend be affected by overfitting. To exclude this, efforts to collect a large Europe-wide dataset of high-quality SSI ground measurements are needed which also include strongly under-represented regions.

Without fine-tuning, the emulator has a moderately higher RMSE (about 6-16 $Wm^{-2}$) than Heliosat SARAH-3, as shown in Figure \ref{fig:scatter_rmse}. Importantly, fine-tuning on any ground station network allows the emulator to outperform SARAH-3 SSI in terms of RMSE with regard to IEA PVPS ground stations. The initial stage of fine-tuning demonstrated performance improvements, as evidenced by reduced RMSEs when compared to the non-fine-tuned emulator across European stations (Table \ref{tab:RMSE}). 
The second and third stages of fine-tuning (on the MeteoSwiss and IEA PVPS stations) achieved a significant reduction in RMSE at nearly all sites of the four ground station networks, as shown in Figure~\ref{fig:scatter_rmse}. 
Remote stations such as in Tunisia, Algeria and Morocco exhibited RMSE reductions of 10--20 $Wm^{-2}$ compared to SARAH-3. 
These improvements in RMSE and relatively stable MBE values suggest that our retrieval models generalize well to diverse geographical locations after fine-tuning on multiple ground station networks. 

In most of Central Europe, the MBE of model E ('Trained on DWD') is low and its RMSE is close to that of the fine-tuned models, but for the stations to the south -- the Spanish and four North African sites -- it shows a bias of 50--100 $\, Wm^{-2}$ (Table \ref{tab:MBE}) and worse accuracy than SARAH-3 at 30--47 $\, Wm^{-2}$ higher RMSE (Table \ref{tab:RMSE}). The model that was fine-tuned on the DWD stations ('Finetuned DWD', model B) did not show these large biases in the North African stations and improved in accuracy compared to the emulator.

\subsubsection{Alpine validation}
The emulator demonstrates a significant enhancement in performance across Alpine ground stations, as presented in Figures \ref{fig:scatter_mbe} and \ref{fig:scatter_rmse}, particularly in regions with large RMSD relative to SARAH-3 (Figure \ref{fig:regional_analysis}).
Some of the ground stations, such as Gornergrat (GOR) at 3129 m a.s.l., are located at high altitudes on mountain ranges. Even though the emulator was trained solely on SARAH-3 SSI without any ground stations data, it achieves RMSE reductions of up to 50 W/m² compared to SARAH-3, indicating an improvement in retrieval accuracy. 

Comparing SSI from Alpine stations such as Gornergrat to Heliosat SARAH-3 reveals a systematic underestimation of SSI by SARAH-3, as indicated by a pronounced off-diagonal frequency peak in Figure \ref{fig:gor_qqplot}(a). This underestimation, which likely results from the misclassification of snow cover as clouds by Heliosat SARAH-3 \citep{carpentieriSatellitederivedSolarRadiation2023}, affects the majority of Alpine stations (not shown). 
Notably, the emulator does not replicate this pattern, even without fine-tuning (Figure \ref{fig:gor_qqplot}(b)), which suggests that the spectral signatures of snow cover differ sufficiently from those of clouds to allow for improved distinction.

Two anomalies in the retrieval of Alpine SSI are evident in Figure \ref{fig:scatter_mbe}, involving models B ('Finetuned on DWD') and E ('Trained on DWD'). Both models were trained on DWD stations and display significant biases at certain Alpine stations. Model B, which was fine-tuned on DWD data, exhibits a bias exceeding 150 W/m² across the entire Alps. 
The scatter plot for model B ('Finetuned on DWD') at GOR (Figure \ref{fig:gor_qqplot}(B)) shows that this bias stems from an overestimation at low SSI values. Conversely, model E ('Trained on DWD') exhibits a substantial negative bias ($<$-150 W/m²) at some Alpine stations, including GOR, as reflected by the underestimation in the scatter plot (Figure \ref{fig:gor_qqplot}(E)). Similar patterns are observed at other high-altitude stations.
One possible explanation for the underestimation with model E is that the spectral signature of Alpine surfaces under clear-sky conditions differs markedly from the surfaces predominant in the DWD training set.


\subsubsection{Representativeness error}
\label{sec:representativeness}
Meteosat SEVIRI Level-1.5 data are provided at a typical grid size of 3$\times$3\,km, so the SSI estimates represent instantaneous values of the mean SSI in a 3$\times$3\,km area. In contrast, SSI ground observations provide point measurements that are averaged over periods of 1-10 minutes and have limited spatial representativeness. \citet{huangRepresentativenessErrorsPointscale2016} evaluated the representative errors for an SSI product on a grid scale of 5$\times$5\,km.
Seventeen pyranometer stations within a 5$\times$5\,km area at a location in Northern China measured subgrid scale deviations. 
At a 10-minute averaging of ground observations, they found the average representativeness error of $0.05^\circ$ grid-scale products to be $13.4\%$ with an average RMSE of $93.2~Wm^{-2}$ for the SSI product. With increasing cloud cover fraction, this error increases to around $32.5\%$ with an RMSE of $113.8~Wm^{-2}$. Clouds increase the representativeness error due to subgrid inhomogeneity, but also solar retrieval products do not take into account 3D radiative transfer effects within clouds \citep{huangRepresentativenessErrorsPointscale2016}.
While representativeness errors may, in principle, affect our validation, we have demonstrated above that fine-tuning our retrieval models of instantaneous SSI on 10-minute averaged ground observations still increases the performance by up to $20~Wm^{-2}$. This shows that, despite potential representativeness errors, our deep-learning retrieval models learn systematic spatial SSI patterns to provide more accurate representative SSI retrieval.

\subsection{Cloudy versus clear-sky conditions}
\label{par:cloudy_vs_clear} 
\begin{figure}
    \centering
    \makebox[\textwidth][c]{\includegraphics[width=1\textwidth]{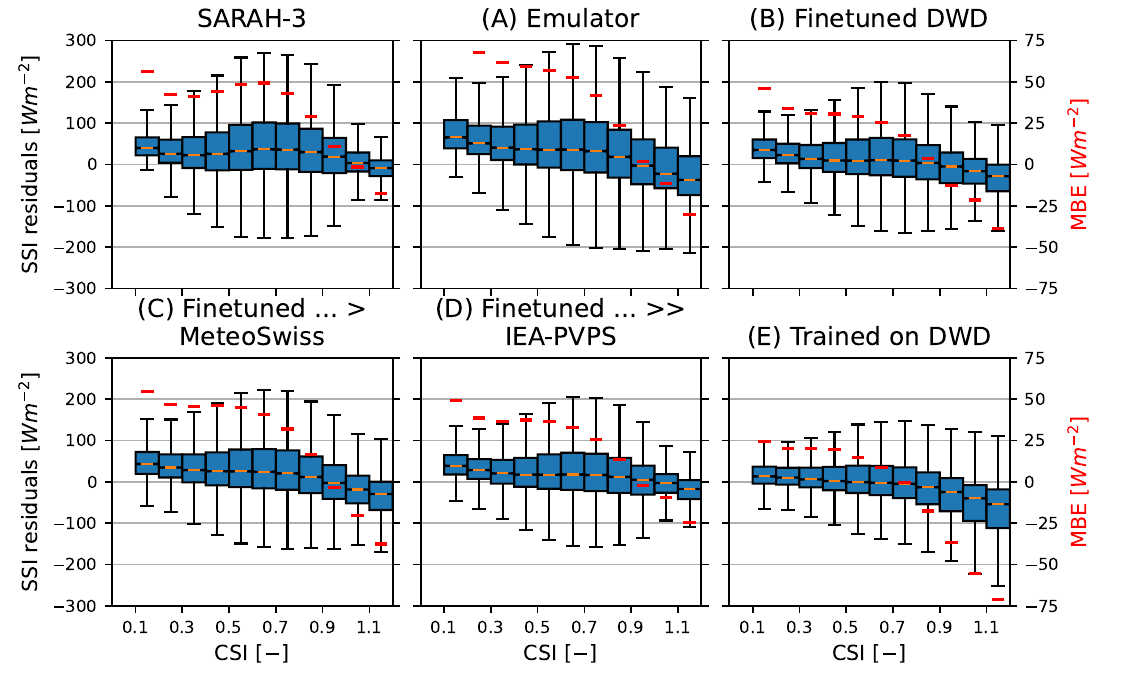}}%
    \caption{Instantaneous SSI residual boxplot on the IEA-PVPS dataset binned on the clear-sky index, $\hat{SSI} - SSI$. The red horizontal line specifies the mean bias error for each bin.}
    \label{fig:CSI_boxplot}
\end{figure}

The performance of our SSI retrieval models is discussed for different cloudiness conditions in the following. SSI retrieval errors depend on the level of cloudiness present. 
\citet{quesada-ruizAdvancedANNbasedMethod2015} showed that the RMSE of Heliosat-2 SSI with regard to ground stations depends on the CSI. 
They presented a retrieval method that achieved the largest reductions in RMSE in overcast conditions, where the Heliosat retrievals had the highest RMSE. 
Likewise, \citet{verboisRetrievalSurfaceSolar2023} accomplished the largest improvements in RMSE for samples with CSI $<$ 0.6.
We investigate the performance of our five retrieval models in clear-sky and cloudy conditions by separating the ground station measurements into CSI classes.
The clear-sky index was calculated from measured SSI at the ground stations and using the Ineichen-Perez clear-sky surface radiation model~\citep{perezModelingDaylightAvailability1990}.
Figure \ref{fig:CSI_boxplot} shows SSI error statistics for different levels of cloudiness as expressed by the CSI. 
Heliosat SARAH-3 underestimates SSI by around -25 $Wm^{-2}$ in clear-sky index larger than 1 and overestimates SSI by around 27--90 $Wm^{-2}$ for CSI $<$ 0.8. 
The large systematic biases in SARAH-3 and other radiative-transfer-based SSI retrievals (positive for CSI $<$ 0.8 and negative for CSI $\geq$ 1) \citep{ineichenSatelliteApplicationFacilities2009} may partially be related to fractional clouds~(J\"org Trentmann, personal communication). 
Clouds that partially cover the grid cell are measured at a point location while the satellite observes an average. 
The clear-sky index defined on the basis of ground observations and not over the entire grid cell can result in these biases.
The emulator and the fine-tuned models mimic the negative and positive biases of Heliosat SARAH-3 (Figure \ref{fig:CSI_boxplot}). 
While the emulator generates somewhat larger residuals, fine-tuning on ground stations effectively decreased the residual spread for CSI $<$ 0.8 compared to SARAH-3. 

Interestingly, solely training on ground stations without any Heliosat emulation (model E 'Trained on DWD') generates the smallest biases in cloudy conditions (CSI $<$ 0.8). 
The model trained only on ground stations (model E) also features the best generalizability to locations that are remote from its training set sites. 
Therefore, in cloudy conditions, the model trained solely on DWD stations provides SSI estimates with the smallest biases and RMSEs even at out-of-domain locations compared to the other models and with large performance improvements over Heliosat (Figure \ref{fig:CSI_boxplot}). 
The superior SSI retrieval performance of model E is exemplified in 
Figure \ref{fig:ouj_qqplot} for the Moroccan IEA-PVPS station. 
Even though the model was trained only on ground stations in Germany without any training data from any other regions, it shows the best SSI retrieval performance in cloudy conditions even in countries such as Morocco that are thousands of kilometers from the training set domain and feature very different climates and surface albedos. 
The model shows the best performance with an improvement of 40 $Wm^{-2}$ in RMSE with respect to SARAH-3. The other North African and Spanish stations exhibit the same effect.
The respective performance in terms of MBE and RMSE on the IEA-PVPS dataset for cloudy CSI $<$ 0.8 and clear-sky CSI $\geq$ 0.8 samples is depicted in Tables \ref{tab:RMSE} and \ref{tab:MBE}.
Models trained on ground stations of DWD and MeteoSwiss show in general a worse RMSE on the North-African and Spanish stations in clear-sky conditions than SARAH-3, while in cloudy conditions the performance improves compared to SARAH-3.
In cloudy conditions, the model trained solely on DWD stations (model E, 'Trained on DWD') has the lowest RMSE and lowest absolute MBE for all IEA-PVPS stations (Figure \ref{fig:CSI_boxplot}), demonstrating the model's generalizability to out-of-domain locations. 

Our results indicate that the generalizability of a data-driven SSI retrieval model is not only related to the model training data or method, but also depends on the level of cloudiness present in the location for which SSI is retrieved with the data-driven model. This observation can provide additional explanation for the insufficient SSI retrieval generalizability seen by \citet{verboisRetrievalSurfaceSolar2023} and \citet{yangWorldwideValidationEarth2022}. 

Based on the validation on ground station observations in section \ref{sec:validation} (Figures \ref{fig:scatter_mbe} and \ref{fig:scatter_rmse}), we find that if a model is trained on ground station SSI data from a given region, it can estimate clear-sky SSI effectively with similar performance as Heliosat for that region, but not necessarily for other regions. In contrast, in cloudy conditions, a model trained on only DWD stations can estimate SSI accurately even in locations far away from the training set domain (out-of-domain generalization), achieving large performance gains compared to Heliosat.
In clear-sky conditions, data-driven SSI retrieval models predict SSI based of reflectivity and emissivity from the surface. 
Different land classes have different backscatter properties in the reflectivity bands and different brightness temperature signatures in the emissivity bands.
The surface characteristics seen by SEVIRI differ largely from region to region and the relationship between observed irradiances and SSI is complex.
It seems that for generalizable retrieval in clear-sky conditions all different surface characteristics need to be represented in the training dataset.
We hypothesize that clouds, on the other hand, have more similar spectral signatures across different latitudes and that, therefore, data-driven SSI retrieval models tend to generalize better to regions outside their training set domain in cloudy conditions than in clear-sky conditions. 

\FloatBarrier

\subsection{Importance of SEVIRI channels for estimating SSI}
\citet{shiFirstEstimationHighresolution2023} showed first the importance of individual channels and attributes to the retrieval of SSI.
As the channels relate to different wavelength bands, they differ in their relative importance for capturing atmospheric composition aspects and surface properties~\citep{schmetzIntroductionMeteosatSecond2002}. 
For example, the visible VIS006 and VIS008 are more relevant for cloud opacity due to reflectivity whereas WV062 and WV073 mainly sense tropospheric water vapor. 
Here we continue the trend of explaining the importance of features but for neural network-type models.
We quantified the relevance of the individual SEVIRI channels in an observations-based, data-driven manner based on the permutation feature importance of the predictor variables of each of our five SSI retrieval models.
Permutation feature importance methods assess the importance of each predictor variable (feature) in a machine learning model based on how much the predictor variable contributes to the model's performance~\citep{fisherAllModelsAre2019a}. 
Permutation feature importance quantifies how the model performance changes when the values of a predictor variable are randomly permuted while keeping the values of the other predictor variables unchanged. 
If the random shuffling leads to a significant drop in model accuracy, the predictor variable whose values were shuffled is important for the model's performance.
After permuting a predictor variable over the entire validation set, we calculated the model performance in terms of the mean absolute error (MAE) and compared it to the model performance without permutation. 
The 15$\times$15 grids represent a single feature per channel in the feature importance.

\begin{figure}[h!]
    \centering
    \includegraphics[width=\textwidth]{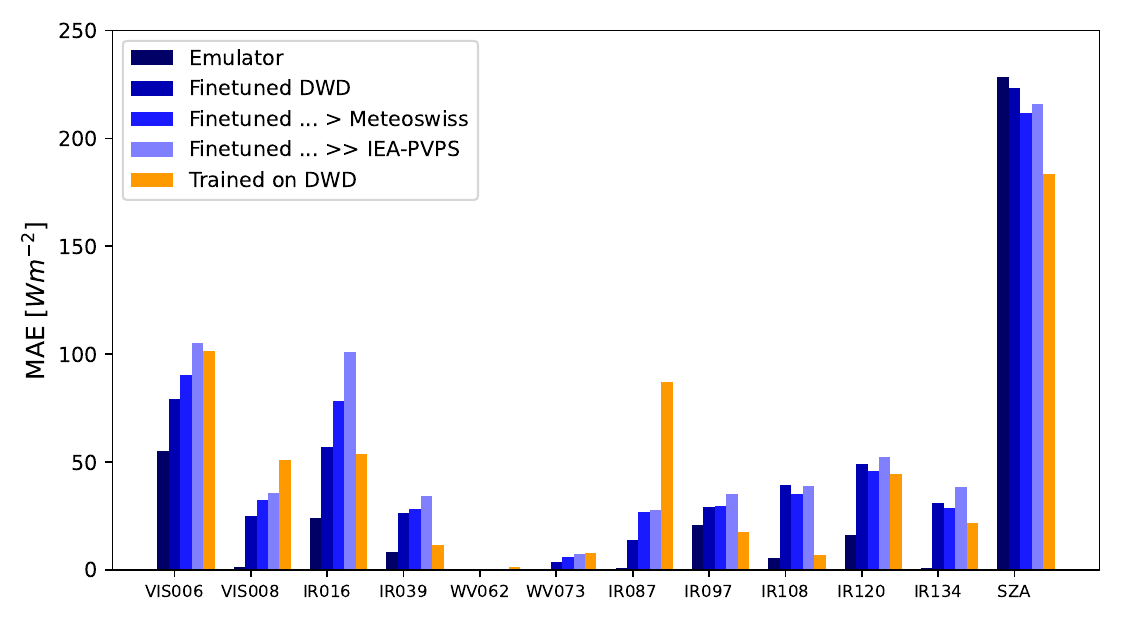}
    \caption{Permutation feature importance in increased instantaneous Mean Absolute Error (MAE) on the IEA-PVPS dataset.}
    \label{fig:feature_importance}
\end{figure}

The permutation feature importance is shown in Figure \ref{fig:feature_importance} based on the IEA-PVPS dataset. Shuffling a predictor variable that does not affect the retrieval model's performance, such as WV062, leaves the MAE practically unchanged. Shuffling features essential to the retrieval model -- such as the VIS006 channel and the SZA -- cause large performance decreases of around 100 and 200 $Wm^{-2}$ in MAE, respectively. The predictor variables latitude, longitude, azimuth, DEM and day of the year were left out from Figure \ref{fig:feature_importance} as permutation feature importance showed they had no significant effect on the model performance.
In \citet{shiFirstEstimationHighresolution2023} the solar azimuth angle shows significant importance. 
The difference might be that the surrounding pixel information given to the CNN might make the solar azimuth angle redundant.
The emulator effectively disregards the VIS008 channel. However, the other deep learning retrieval models retrieve information and gain performance from the VIS008 channel. The emulator shows the lowest feature importance for all channels but attributes the largest importance to the SZA, compared to the other models, which suggests it does not yet make optimal use of the information contained in the SEVIRI channels.

Figures \ref{fig:feature_importance_clearsky} and \ref{fig:feature_importance_cloudy} illustrate predictor importance for clear-sky and cloudy conditions. 
As expected, our SSI retrieval models rely mostly on SZA in clear-sky conditions, because the SZA includes the majority of information to estimate SSI without cloud attenuation. 
In cloudy conditions, the visible, near-infrared and infrared channels gain importance. 
For model E ('Trained on DWD'), the IR087 channel plays a significant role in the clear-sky conditions compared to the other retrieval models. 
IR087, and a combination of IR108 and IR120, are important for distinguishing ice clouds from water clouds \citep{schmetzIntroductionMeteosatSecond2002} and may be helpful in distinguishing snow cover from clouds in clear-sky conditions.
We expect that a data-driven evaluation of channel importance can contribute to further improvement of radiative-transfer based models.

\FloatBarrier

\section{Conclusions}
\label{sec:conclusion}

Our study introduced a machine-learning-based satellite retrieval for instantaneous surface solar irradiance and demonstrated its capability to provide accurate and generalizable SSI estimates across Europe. Our retrieval models were trained on multiple years of Meteosat SEVIRI observations in eleven visible and infrared channels. Heliosat SSI estimates and SSI from ground station networks were used as target variables in the machine learning model development. Instantaneous SSI is not averaged in time and primarily of interest in short-term forecasting, such as intra-hour solar forecasts for applications in solar energy production and building control. Our study is the first to emulate a radiative-transfer based retrieval model. We introduced a deep learning emulator of a Heliosat SSI retrieval (SARAH-3) and showed that the emulator is an accurate surrogate of thereof. By including SSI measurements from ground stations in the model training, our SSI retrieval model outperformed Heliosat in terms of accuracy and biases in cloudy conditions (CSI $<$ 0.8). When trained on ground station SSI, our retrieval models generalized well to regions with other climates and surface albedos. Our results indicate that the generalizability of a data-driven SSI retrieval model is not only related to the model training data or training method, but also depends on the amount of cloudiness present in the location at which SSI is retrieved with the data-driven model. We found that, in cloudy conditions, a model trained only on ground stations can estimate SSI accurately even in locations with different surface characteristics, far away from the training test domain. 
Our SSI retrieval model trained on ground stations provides out-of-domain generalization and achieves large performance gains compared to Heliosat. Moreover, we found emulation is helpful but not necessary to achieve good generalizability. In fact, for cloudy conditions (CSI $<$ 0.8), better generalizability (the lowest bias and RMSE) was achieved by direct model training on ground stations.
While transfer learning did not increase performance for cloudy conditions, it did help with spatial generalization in clear-sky conditions where the model trained solely on ground stations predicted SSI with large biases. 

We also showed that the SSI of Heliosat SARAH-3 exhibits large biases in mountain regions, and that training and fine-tuning our retrieval models on ground station SSI strongly reduces these large biases, outperforming Heliosat. Further, our study quantified the relevance of the Meteosat channels and other predictor variables, such as solar zenith angle and time of year, for the model accuracy in different cloud conditions. Permutation feature importance showed that in cloudy conditions, multiple near-infrared and infrared channels enhance performance. Our results can facilitate the development of more accurate radiative-transfer-based SSI retrieval models.
Future research should investigate additional predictor variables and their impact on the retrieval quality, notably to improve the performance in clear-sky conditions or to replace the clear-sky retrieval with a physical model to create a hybrid model.

Our study shows it is possible to emulate a RTM-based retrieval model with DL. 
Finetuning this base model on ground stations will lead to a better generalizable model with high performance but there are limitations and difficulties to the methodology. 
Finetuning requires quite a bit of trial and error and advanced training methods from the field of transfer learning to not overfit the new type of dataset. 
Unforeseen biases in particular climates, such as mountainous regions of the domain, can still occur using a finetuned model.
The general accuracy gain of up to 20\% compared to SARAH-3 in cloudy samples should prove the usefulness of DL retrieval since those are most important for short-term solar forecasts.
We expect our contribution will facilitate faster, better and more generalizable SSI retrieval with better knowledge of the pitfalls DL retrieval brings.

\section{Acknowledgements}
We thank the anonymous reviewers for the constructive comments.
This work was supported by the Swiss Innovation Agency Innosuisse through grant 104.531 IP-EE. We acknowledge helpful discussions with J\"org Trentmann and Uwe Pfeifroth of DWD. We made use of the pyranometer measurements from IEA-PVPS that are accessible via ~\citep{forstinger_2023_7867003}. 
The DWD and KNMI pyranometer measurements are available from \citet{dwd10minuteStationObservations2024} and \citet{knmiSunshineRadiation102024}. The MeteoSwiss pyranometer data were provided by the Swiss Federal Office of Meteorology MeteoSwiss via IDAWEB~\citep{meteoswissGlobalRadiationTen2024}. The digital elevation model was provided by Copernicus~\citep{airbusCopernicusDigitalElevation2024}.

\section{Declaration of generative AI and AI-assisted technologies in the writing process.}
During the preparation of this work, the first author used ChatGPT in order to help with writing style. After using this tool, the author reviewed and edited the content as needed and takes full responsibility for the content of the published article.

\clearpage

\bibliographystyle{tfcad}
\bibliography{SIE}

\newpage

\appendix
\section{Architecture}
\label{app:trainingdetails}

\begin{figure}[h]
    \centering
    \makebox[\textwidth][c]{\includegraphics[width=1.\textwidth]{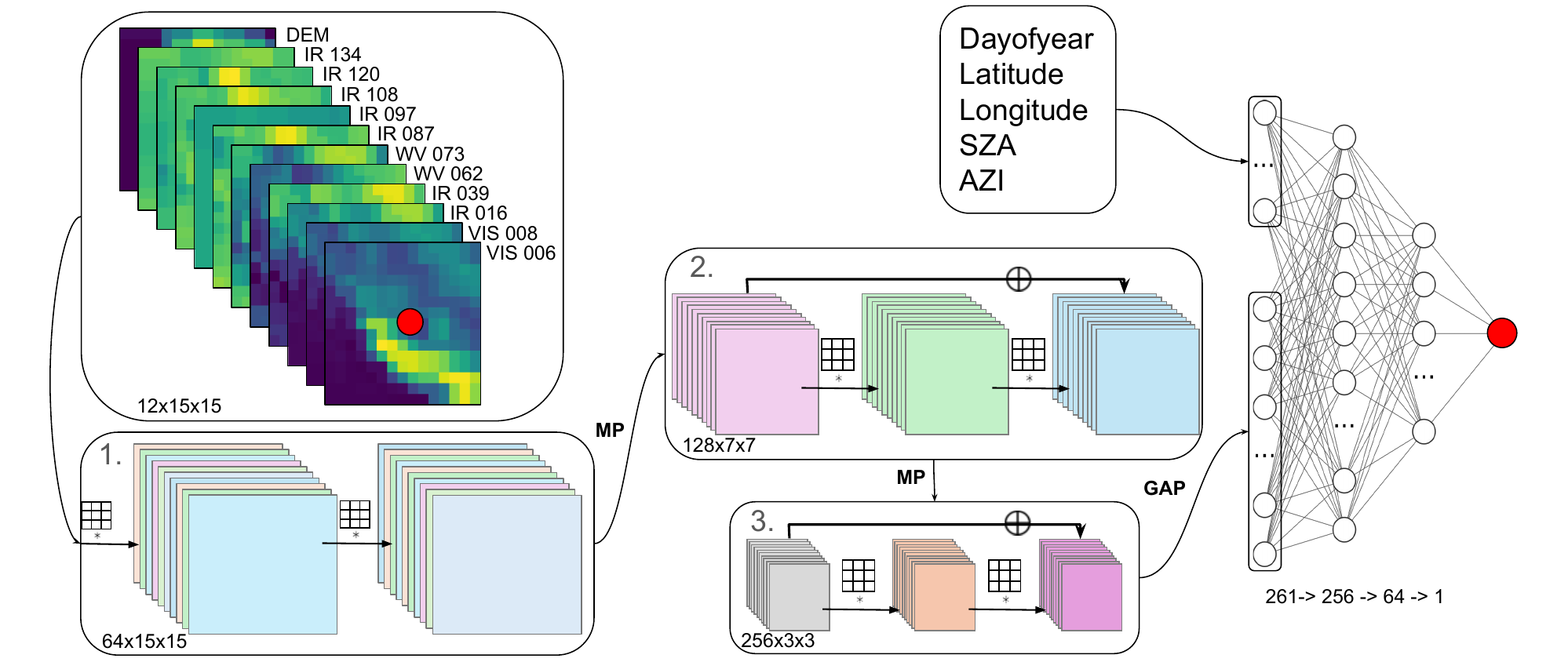}}
    \caption{The architecture of the convolutional residual network. MP stands for 2$\times$2 max pooling and GAP for global average pooling in the spatial dimensions.}
    \label{fig:convresnet}
\end{figure}

\newpage

\section{Cloudy and Clear-sky comparisons}

\begin{table}[h!]
    \tbl{Instantaneous Root Mean Square Error (RMSE) [$Wm^{-2}$] calculated on all IEA-PVPS stations in the dataset. Clear-sky and cloudy are defined as $CSI \geq 0.8$ and CSI $<$ 0.8.}{
    \input{RMSE_table}
    }
    \label{tab:RMSE}
\end{table}

\begin{table}[h!]
    \tbl{Instantaneous Mean Bias Error (MBE) [$Wm^{-2}$] calculated on all IEA-PVPS stations in the dataset. Clear-sky and cloudy are defined as CSI $\geq$ 0.8 and CSI $<$ 0.8.}{
    \input{MBE_table}}
    \label{tab:MBE}
\end{table}

\begin{figure}[h!]
    \centering
    \includegraphics[width=\textwidth]{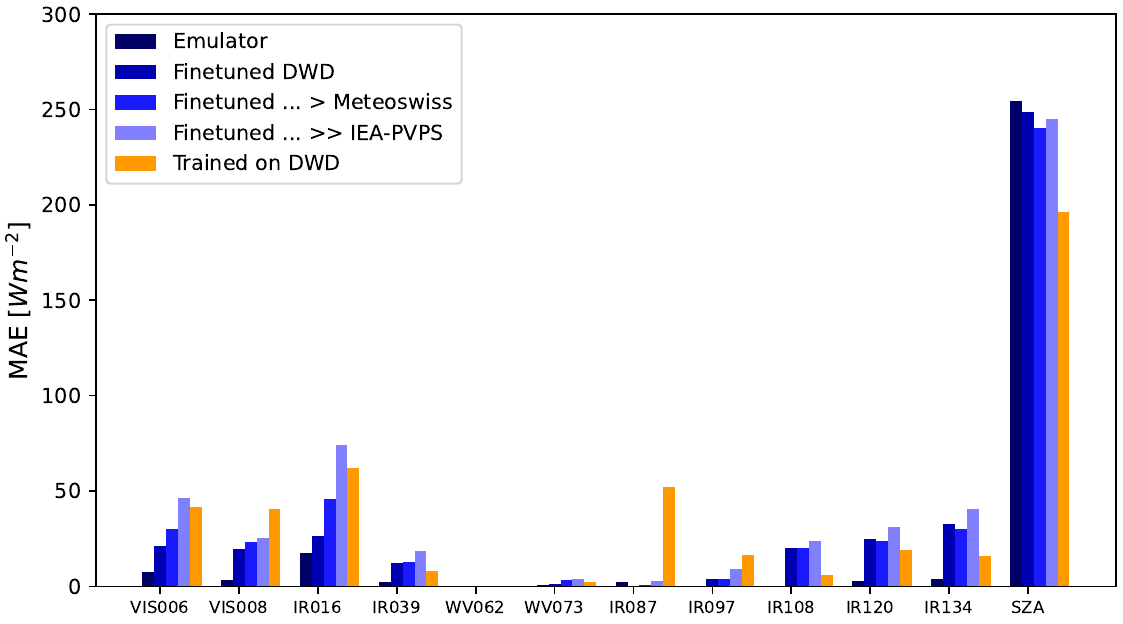}
    \caption{Permutation feature importance in clear-sky conditions (CSI $\geq$ 0.8) in instantaneous Mean Bias Error (MBE) concerning the IEA-PVPS dataset.}
    \label{fig:feature_importance_clearsky}
\end{figure}

\begin{figure}[h!]
    \centering
    \includegraphics[width=\textwidth]{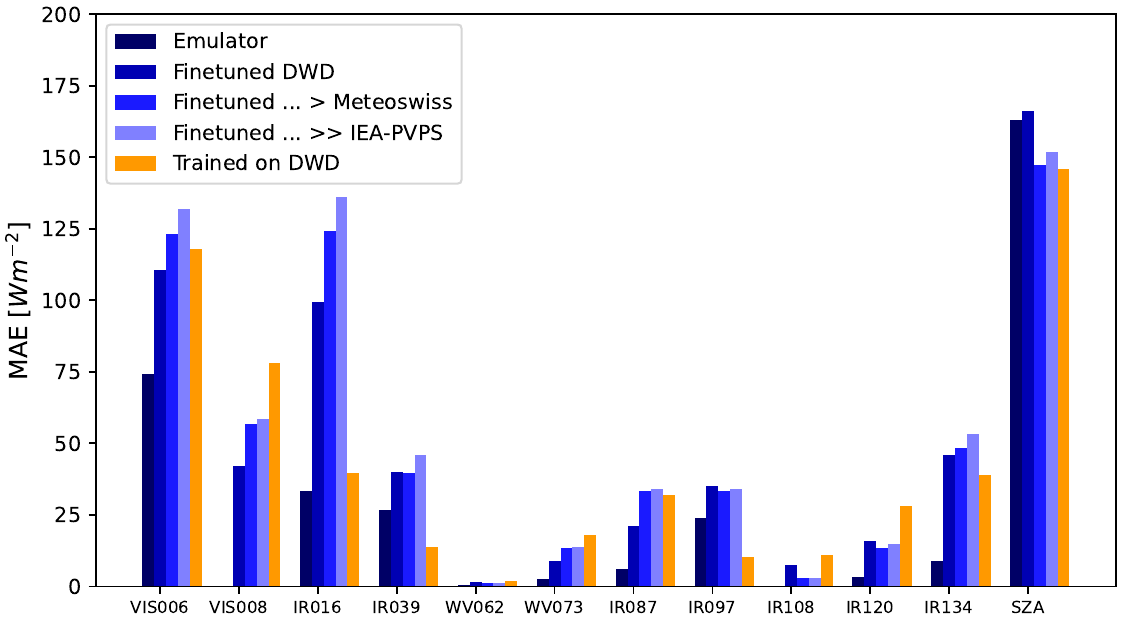}
    \caption{Permutation feature importance in cloudy conditions (CSI $<$ 0.8) in instantaneous Mean Bias Error (MBE) concerning the IEA-PVPS dataset.}
    \label{fig:feature_importance_cloudy}
\end{figure}

\FloatBarrier
\section{Scatter plots}

\begin{figure}[h!]
\centering
\includegraphics[width=.95\textwidth]{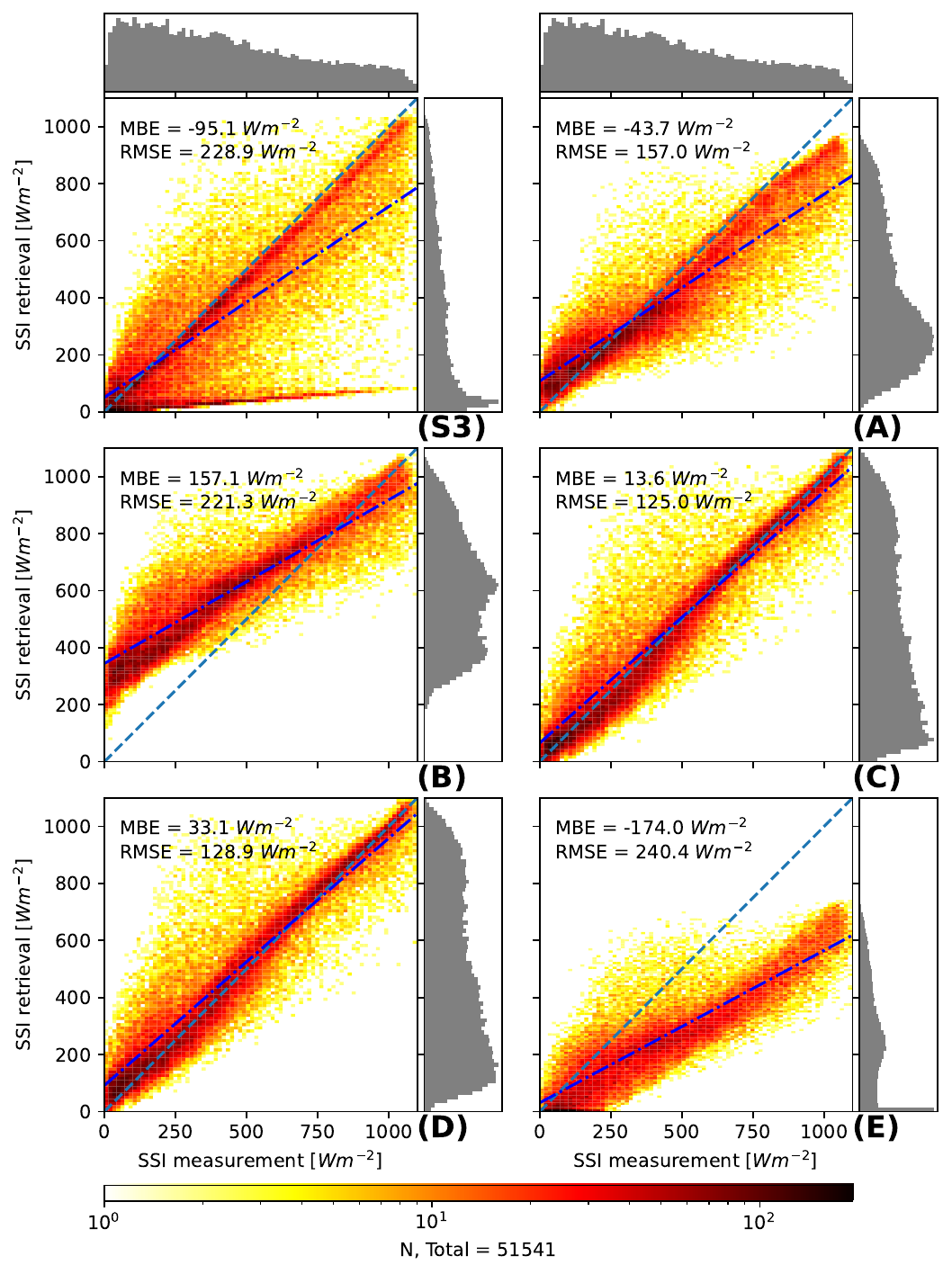}
\caption{Scatter plot of instantaneous SSI retrieval for Alpine station Gornergrat at 3012~m (S3) SARAH-3, (A) Emulator, (B) Finetuned on DWD, (C) Finetuned ...$>$ MeteoSwiss (D) Finetuned ...$>>$ IEA-PVPS (E) Trained on DWD.}
\label{fig:gor_qqplot}
\end{figure}

\begin{figure}[h!]
\centering
\includegraphics[width=.95\textwidth]{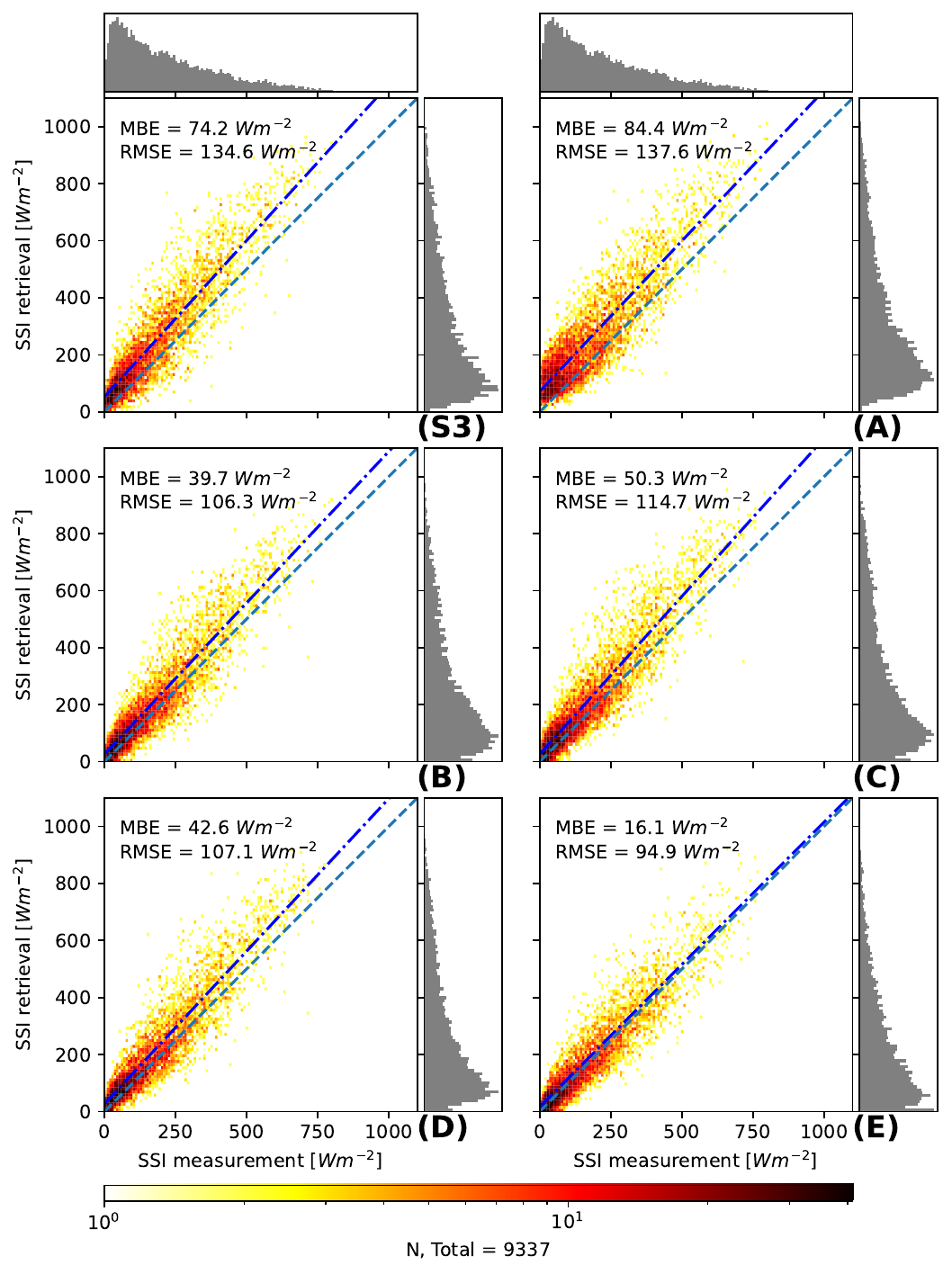}
\caption{Scatter plot of instantaneous SSI retrieval for IEA-PVPS station Oujda (OUJ) filtered on CSI $<$ 0.8: (S3) SARAH-3, (A) Emulator, (B) Finetuned on DWD, (C) Finetuned ... $>$ MeteoSwiss (D) Finetuned ...$>>$ IEA-PVPS (E) Trained on DWD.}
\label{fig:ouj_qqplot}
\end{figure}

\end{document}

%% file: models_table.tex
\begin{tabular}{llccccccccc}
\toprule
& & \multicolumn{5}{c}{Target data seen}  & \multirow{2}{*}{\rotatebox{90}{Learning rate}}  & \multicolumn{2}{c}{\begin{tabular}[c]{@{}c@{}}Proximal\\ 
    loss\end{tabular}}  &  \\ \cline{3-7} \cline{9-10} 
\multicolumn{2}{c}{Models} & \multicolumn{1}{c}{\rotatebox{90}{SARAH-3}} & \multicolumn{1}{c}{\rotatebox{90}{DWD}} & \multicolumn{1}{c}{\rotatebox{90}{MeteoSwiss}} & \multicolumn{1}{c}{\rotatebox{90}{IEA-PVPS}} & \rotatebox{90}{KNMI} &                                                                & \multicolumn{1}{c}{\begin{tabular}[c]{@{}c@{}} Weights \\ $\theta_0$\end{tabular}}   & $\alpha$ & FCN freeze  \\ \hline
A & Emulator   & \multicolumn{1}{c}{x}       & \multicolumn{1}{c}{}     & \multicolumn{1}{c}{}             & \multicolumn{1}{c}{}         &      & \begin{tabular}[c]{@{}c@{}}$10^{-4}$ \\ - $10^{-5}$\end{tabular} & \multicolumn{1}{c}{-}                                                                                 & -        & -                                                                      \\ \hline
B & \begin{tabular}[c]{@{}l@{}}Finetuned \\ DWD\end{tabular}                                       & \multicolumn{1}{c}{x}       & \multicolumn{1}{c}{x}    & \multicolumn{1}{c}{}             & \multicolumn{1}{c}{}         &      & $10^{-6}$                                                      & \multicolumn{1}{c}{Emulator}                                                                          & 5000     & Yes                                                                    \\ \hline
C & \begin{tabular}[c]{@{}l@{}}Finetuned \\ ...\textgreater\\ MeteoSwiss\end{tabular}              & \multicolumn{1}{c}{x}       & \multicolumn{1}{c}{x}    & \multicolumn{1}{c}{x}            & \multicolumn{1}{c}{}         &      & $10^{-6}$                                                      & \multicolumn{1}{c}{\begin{tabular}[c]{@{}c@{}}Finetuned \\ DWD\end{tabular}}                          & 10000    & Yes                                                                    \\ \hline
D & \begin{tabular}[c]{@{}l@{}}Finetuned \\ ...\textgreater{}\textgreater \\ IEA-PVPS\end{tabular} & \multicolumn{1}{c}{x}       & \multicolumn{1}{c}{x}    & \multicolumn{1}{c}{x}            & \multicolumn{1}{c}{x}        &      & $10^{-7}$                                                      & \multicolumn{1}{c}{\begin{tabular}[c]{@{}c@{}}Finetuned \\ ...\textgreater\\ MeteoSwiss\end{tabular}} & 10000    & No                                                                     \\ \hline \hline
E & \begin{tabular}[c]{@{}l@{}}Trained on \\ DWD\end{tabular}                                      & \multicolumn{1}{c}{}        & \multicolumn{1}{c}{x}    & \multicolumn{1}{c}{}             & \multicolumn{1}{c}{}         &      & $10^{-5}$                                                      & \multicolumn{1}{c}{-}                                                                                 & -        & -                                                                      \\ \bottomrule
\end{tabular}

%% file: RMSE_table.tex
\begingroup
\renewcommand{\arraystretch}{.90}
\setlength{\tabcolsep}{6pt}

\begin{tabular}{lC{1.1cm}C{1.1cm}|C{.8cm}C{.8cm}C{.8cm}C{.8cm}C{.8cm}C{.8cm}} \toprule

RMSE & \centering \rotatebox{90}{Station} & \rotatebox{90}{Samples}& \rotatebox{90}{SARAH-3} & \rotatebox{90}{ (A) Emulator} & \rotatebox{90}{\makecell[c]{(B) Finetuned\\DWD}} & \rotatebox{90}{\makecell[c]{(C) Finetuned $>$\\MeteoSwiss}} & \rotatebox{90}{\makecell[c]{(D) Finetuned $>>$\\IEA-PVPS}} & \rotatebox{90}{\makecell[c]{(E) Trained\\on DWD}} \\ \hline
 & CAB & 27183 & 82 & 93 & 73 & 79 & 75 & \cellcolor[HTML]{C0C0C0}\textbf{69} \\  
 & CAR & 23488 & 81 & 87 & 68 & 69 & \cellcolor[HTML]{C0C0C0}\textbf{65} & 71 \\  
 & CEN & 24431 & 96 & 101 & 83 & 85 & \cellcolor[HTML]{C0C0C0}\textbf{81} & 83 \\  
 & GHA & 13496 & 87 & 101 & 92 & 97 & \cellcolor[HTML]{C0C0C0}\textbf{72} & 141 \\  
 & MIL & 21977 & 78 & 87 & 67 & 68 & 65 & \cellcolor[HTML]{C0C0C0}\textbf{64} \\  
 & MIS & 15767 & 91 & 99 & 83 & 91 & \cellcolor[HTML]{C0C0C0}\textbf{79} & 110 \\  
 & NOR & 29248 & 82 & 91 & 74 & 75 & 73 & \cellcolor[HTML]{C0C0C0}\textbf{69} \\  
 & OUJ & 35691 & 85 & 98 & 80 & 83 & \cellcolor[HTML]{C0C0C0}\textbf{74} & 86 \\  
 & PAL & 30438 & 89 & 98 & 81 & 81 & \cellcolor[HTML]{C0C0C0}\textbf{79} & 80 \\  
 & PAY & 28228 & 87 & 97 & 74 & 78 & 74 & \cellcolor[HTML]{C0C0C0}\textbf{70} \\  
 & TAB & 40207 & 81 & 93 & 81 & 85 & \cellcolor[HTML]{C0C0C0}\textbf{74} & 115 \\  
 & TAT & 25627 & 79 & 96 & 91 & 89 & \cellcolor[HTML]{C0C0C0}\textbf{64} & 125 \\  
 & TOR & 28262 & 86 & 96 & 78 & 83 & 78 & \cellcolor[HTML]{C0C0C0}\textbf{74} \\  
\multirow{-14}{*}{All} & VIS & 26770 & 67 & 80 & 63 & 67 & 62 & \cellcolor[HTML]{C0C0C0}\textbf{59} \\ \hline 
 & CAB & 46\% & 82 & 95 & 69 & \cellcolor[HTML]{C0C0C0}\textbf{60} & 62 & 75 \\  
 & CAR & 68\% & 63 & 77 & 62 & 60 & \cellcolor[HTML]{C0C0C0}\textbf{55} & 73 \\  
 & CEN & 56\% & 80 & 90 & 83 & 72 & \cellcolor[HTML]{C0C0C0}\textbf{71} & 88 \\  
 & GHA & 85\% & 55 & 83 & 84 & 90 & \cellcolor[HTML]{C0C0C0}\textbf{52} & 146 \\  
 & MIL & 63\% & 62 & 79 & 62 & 54 & \cellcolor[HTML]{C0C0C0}\textbf{54} & 61 \\  
 & MIS & 73\% & 63 & 81 & 65 & 81 & \cellcolor[HTML]{C0C0C0}\textbf{61} & 113 \\  
 & NOR & 47\% & 84 & 98 & 79 & \cellcolor[HTML]{C0C0C0}\textbf{71} & 74 & 74 \\  
 & OUJ & 74\% & \cellcolor[HTML]{C0C0C0}\textbf{58} & 79 & 69 & 68 & 59 & 83 \\  
 & PAL & 47\% & 85 & 98 & 91 & \cellcolor[HTML]{C0C0C0}\textbf{77} & 80 & 92 \\  
 & PAY & 53\% & 81 & 88 & 71 & \cellcolor[HTML]{C0C0C0}\textbf{61} & 62 & 70 \\  
 & TAB & 78\% & \cellcolor[HTML]{C0C0C0}\textbf{59} & 84 & 74 & 78 & 63 & 122 \\  
 & TAT & 83\% & 55 & 80 & 90 & 86 & \cellcolor[HTML]{C0C0C0}\textbf{52} & 132 \\  
 & TOR & 43\% & 85 & 102 & 77 & \cellcolor[HTML]{C0C0C0}\textbf{72} & 73 & 72 \\  
\multirow{-14}{*}{Clear-sky} & VIS & 53\% & 65 & 75 & 55 & 55 & 56 & \cellcolor[HTML]{C0C0C0}\textbf{54} \\ \hline 
 & CAB & 54\% & 83 & 92 & 76 & 92 & 84 & \cellcolor[HTML]{C0C0C0}\textbf{62} \\  
 & CAR & 32\% & 108 & 106 & 78 & 85 & 81 & \cellcolor[HTML]{C0C0C0}\textbf{68} \\  
 & CEN & 44\% & 112 & 114 & 84 & 98 & 92 & \cellcolor[HTML]{C0C0C0}\textbf{75} \\  
 & GHA & 15\% & 182 & 171 & 129 & 130 & 137 & \cellcolor[HTML]{C0C0C0}\textbf{111} \\  
 & MIL & 37\% & 99 & 99 & 75 & 87 & 81 & \cellcolor[HTML]{C0C0C0}\textbf{69} \\  
 & MIS & 27\% & 141 & 136 & 120 & 114 & 115 & \cellcolor[HTML]{C0C0C0}\textbf{101} \\  
 & NOR & 53\% & 80 & 83 & 69 & 78 & 72 & \cellcolor[HTML]{C0C0C0}\textbf{65} \\  
 & OUJ & 26\% & 135 & 138 & 106 & 115 & 107 & \cellcolor[HTML]{C0C0C0}\textbf{95} \\  
 & PAL & 53\% & 93 & 97 & 71 & 84 & 77 & \cellcolor[HTML]{C0C0C0}\textbf{68} \\  
 & PAY & 47\% & 94 & 107 & 77 & 93 & 85 & \cellcolor[HTML]{C0C0C0}\textbf{71} \\  
 & TAB & 22\% & 128 & 119 & 101 & 103 & 105 & \cellcolor[HTML]{C0C0C0}\textbf{87} \\  
 & TAT & 17\% & 146 & 148 & 95 & 101 & 103 & \cellcolor[HTML]{C0C0C0}\textbf{82} \\  
 & TOR & 57\% & 87 & 91 & 79 & 91 & 82 & \cellcolor[HTML]{C0C0C0}\textbf{75} \\  
\multirow{-14}{*}{Cloudy} & VIS & 47\% & 69 & 85 & 71 & 79 & 69 & \cellcolor[HTML]{C0C0C0}\textbf{64} \\ \bottomrule
\end{tabular}

\endgroup

%% file: MBE_table.tex
\begingroup
\renewcommand{\arraystretch}{.90}
\setlength{\tabcolsep}{6pt}
\begin{tabular}{lC{1.1cm}|C{.8cm}C{.8cm}C{.8cm}C{.8cm}C{.8cm}C{.8cm}}
\toprule
MBE & \rotatebox{90}{Station} & \rotatebox{90}{SARAH-3} & \rotatebox{90}{ (A) Emulator} & \rotatebox{90}{\makecell[c]{(B) Finetuned\\DWD}} & \rotatebox{90}{\makecell[c]{(C) Finetuned $>$\\MeteoSwiss}} & \rotatebox{90}{\makecell[c]{(D) Finetuned $>>$\\IEA-PVPS}} & \rotatebox{90}{\makecell[c]{(E) Trained\\on DWD}} \\ \hline
 & CAB & \cellcolor[HTML]{FEFCFB}2 & \cellcolor[HTML]{FDF6F5}5 & \cellcolor[HTML]{FDF6F5}5 & \cellcolor[HTML]{F4D6D0}22 & \cellcolor[HTML]{F7E1DD}16 & \cellcolor[HTML]{ECF2FC}-10  \\ 
 & CAR & \cellcolor[HTML]{FAEBE8}11 & \cellcolor[HTML]{FEFAF9}3 & \cellcolor[HTML]{EBF1FC}-11 & \cellcolor[HTML]{E9F0FC}-12 & \cellcolor[HTML]{F2F6FD}-7 & \cellcolor[HTML]{C5D8F7}-32 \\ 
 & CEN & \cellcolor[HTML]{F9E9E5}12 & \cellcolor[HTML]{FBEEEC}9 & \cellcolor[HTML]{E7EFFC}-13 & \cellcolor[HTML]{FDFDFE}-1 & \cellcolor[HTML]{FFFEFD}1 & \cellcolor[HTML]{D1E0F9}-25 \\  
 & GHA & \cellcolor[HTML]{F5D8D2}21 & \cellcolor[HTML]{F0C5BC}31 & \cellcolor[HTML]{C8DAF8}-30 & \cellcolor[HTML]{CEDEF8}-27 & \cellcolor[HTML]{FEFCFB}2 & \cellcolor[HTML]{4A86E8}-101 \\  
 & MIL & \cellcolor[HTML]{F8E3DF}15 & \cellcolor[HTML]{FBF0EE}8 & \cellcolor[HTML]{F0F5FD}-8 & \cellcolor[HTML]{FEFCFB}2 & \cellcolor[HTML]{FFFEFD}1 & \cellcolor[HTML]{E9F0FC}-12 \\  
 & MIS & \cellcolor[HTML]{FDF6F5}5 & \cellcolor[HTML]{F8E3DF}15 & \cellcolor[HTML]{FCF2F0}7 & \cellcolor[HTML]{D1E0F9}-25 & \cellcolor[HTML]{F7FAFE}-4 & \cellcolor[HTML]{92B6F1}-60 \\  
 & NOR & \cellcolor[HTML]{F4F7FD}-6 & \cellcolor[HTML]{F5F8FD}-5 & \cellcolor[HTML]{ECF2FC}-10 & \cellcolor[HTML]{FFFEFD}1 & \cellcolor[HTML]{F5F8FD}-5 & \cellcolor[HTML]{EEF4FC}-9 \\  
 & OUJ & \cellcolor[HTML]{F8E3DF}15 & \cellcolor[HTML]{F7E1DD}16 & \cellcolor[HTML]{FDFDFE}-1 & \cellcolor[HTML]{FFFEFD}1 & \cellcolor[HTML]{F9FBFE}-3 & \cellcolor[HTML]{C1D5F7}-34 \\  
 & PAL & \cellcolor[HTML]{FDF6F5}5 & \cellcolor[HTML]{FCF2F0}7 & \cellcolor[HTML]{DEE9FA}-18 & \cellcolor[HTML]{FDFDFE}-1 & \cellcolor[HTML]{F0F5FD}-8 & \cellcolor[HTML]{CFDFF9}-26 \\  
 & PAY & \cellcolor[HTML]{FEFCFB}2 & \cellcolor[HTML]{FAECEA}10 & \cellcolor[HTML]{FFFFFF}0 & \cellcolor[HTML]{F8E3DF}15 & \cellcolor[HTML]{F9E9E5}12 & \cellcolor[HTML]{FBFCFE}-2 \\  
 & TAB & \cellcolor[HTML]{FFFEFD}1 & \cellcolor[HTML]{ECF2FC}-10 & \cellcolor[HTML]{CCDDF8}-28 & \cellcolor[HTML]{BDD3F6}-36 & \cellcolor[HTML]{E5EEFB}-14 & \cellcolor[HTML]{79A5ED}-74 \\  
 & TAT & \cellcolor[HTML]{F5D8D2}21 & \cellcolor[HTML]{F1C8C0}29 & \cellcolor[HTML]{BDD3F6}-36 & \cellcolor[HTML]{D8E5FA}-21 & \cellcolor[HTML]{F9FBFE}-3 & \cellcolor[HTML]{5D93EA}-89 \\  
 & TOR & \cellcolor[HTML]{FDFDFE}-1 & \cellcolor[HTML]{FFFEFD}1 & \cellcolor[HTML]{FFFEFD}1 & \cellcolor[HTML]{F8E5E1}14 & \cellcolor[HTML]{FDF6F5}5 & \cellcolor[HTML]{FEFAF9}3 \\  
 \multirow{-14}{*}{All}  & VIS & \cellcolor[HTML]{ECF2FC}-10 & \cellcolor[HTML]{FDFDFE}-1 & \cellcolor[HTML]{FFFEFD}1 & \cellcolor[HTML]{FAEBE8}11 & \cellcolor[HTML]{FDFDFE}-1 & \cellcolor[HTML]{FEFCFB}2 \\ \hline 
& CAB & \cellcolor[HTML]{CCDDF8}-28 & \cellcolor[HTML]{BAD1F6}-38 & \cellcolor[HTML]{D7E4F9}-22 & \cellcolor[HTML]{EEF4FC}-9 & \cellcolor[HTML]{F0F5FD}-8 & \cellcolor[HTML]{BCD2F6}-37 \\  
 & CAR & \cellcolor[HTML]{F0F5FD}-8 & \cellcolor[HTML]{D7E4F9}-22 & \cellcolor[HTML]{CEDEF8}-27 & \cellcolor[HTML]{C3D7F7}-33 & \cellcolor[HTML]{D5E3F9}-23 & \cellcolor[HTML]{A8C4F3}-48 \\  
 & CEN & \cellcolor[HTML]{DCE8FA}-19 & \cellcolor[HTML]{C5D8F7}-32 & \cellcolor[HTML]{B2CCF5}-42 & \cellcolor[HTML]{BCD2F6}-37 & \cellcolor[HTML]{CADBF8}-29 & \cellcolor[HTML]{9FBEF2}-53 \\  
 & GHA & \cellcolor[HTML]{FDF8F7}4 & \cellcolor[HTML]{F6DBD6}19 & \cellcolor[HTML]{B1CAF5}-43 & \cellcolor[HTML]{B4CDF5}-41 & \cellcolor[HTML]{F2F6FD}-7 & \cellcolor[HTML]{4A86E8}-116 \\  
 & MIL & \cellcolor[HTML]{F5F8FD}-5 & \cellcolor[HTML]{DAE6FA}-20 & \cellcolor[HTML]{CCDDF8}-28 & \cellcolor[HTML]{D5E3F9}-23 & \cellcolor[HTML]{DCE8FA}-19 & \cellcolor[HTML]{CCDDF8}-28 \\  
 & MIS & \cellcolor[HTML]{DAE6FA}-20 & \cellcolor[HTML]{F2F6FD}-7 & \cellcolor[HTML]{F0F5FD}-8 & \cellcolor[HTML]{A4C2F3}-50 & \cellcolor[HTML]{D7E4F9}-22 & \cellcolor[HTML]{6E9EEC}-80 \\  
 & NOR & \cellcolor[HTML]{BCD2F6}-37 & \cellcolor[HTML]{ADC8F4}-45 & \cellcolor[HTML]{B1CAF5}-43 & \cellcolor[HTML]{C3D7F7}-33 & \cellcolor[HTML]{BCD2F6}-37 & \cellcolor[HTML]{BAD1F6}-38 \\  
 & OUJ & \cellcolor[HTML]{F2F6FD}-7 & \cellcolor[HTML]{F0F5FD}-8 & \cellcolor[HTML]{E2EBFB}-16 & \cellcolor[HTML]{E0EAFB}-17 & \cellcolor[HTML]{DCE8FA}-19 & \cellcolor[HTML]{A2C1F3}-51 \\  
 & PAL & \cellcolor[HTML]{C8DAF8}-30 & \cellcolor[HTML]{B4CDF5}-41 & \cellcolor[HTML]{99BBF2}-56 & \cellcolor[HTML]{AFC9F4}-44 & \cellcolor[HTML]{AFC9F4}-44 & \cellcolor[HTML]{94B7F1}-59 \\  
 & PAY & \cellcolor[HTML]{CFDFF9}-26 & \cellcolor[HTML]{C5D8F7}-32 & \cellcolor[HTML]{CFDFF9}-26 & \cellcolor[HTML]{E7EFFC}-13 & \cellcolor[HTML]{EBF1FC}-11 & \cellcolor[HTML]{D5E3F9}-23 \\  
 & TAB & \cellcolor[HTML]{DEE9FA}-18 & \cellcolor[HTML]{C5D8F7}-32 & \cellcolor[HTML]{AFC9F4}-44 & \cellcolor[HTML]{9DBDF2}-54 & \cellcolor[HTML]{CCDDF8}-28 & \cellcolor[HTML]{518AE8}-96 \\  
 & TAT & \cellcolor[HTML]{FCF2F0}7 & \cellcolor[HTML]{F7DFDA}17 & \cellcolor[HTML]{A6C3F3}-49 & \cellcolor[HTML]{C1D5F7}-34 & \cellcolor[HTML]{EBF1FC}-11 & \cellcolor[HTML]{4A86E8}-103 \\  
 & TOR & \cellcolor[HTML]{BCD2F6}-37 & \cellcolor[HTML]{A8C4F3}-48 & \cellcolor[HTML]{BFD4F6}-35 & \cellcolor[HTML]{D5E3F9}-23 & \cellcolor[HTML]{C8DAF8}-30 & \cellcolor[HTML]{CFDFF9}-26 \\  
 \multirow{-14}{*}{Clear-sky} & VIS & \cellcolor[HTML]{BFD4F6}-35 & \cellcolor[HTML]{C1D5F7}-34 & \cellcolor[HTML]{DCE8FA}-19 & \cellcolor[HTML]{F2F6FD}-7 & \cellcolor[HTML]{D8E5FA}-21 & \cellcolor[HTML]{E5EEFB}-14 \\ \hline
  & CAB & \cellcolor[HTML]{F2CEC7}26 & \cellcolor[HTML]{EAB0A4}42 & \cellcolor[HTML]{F2CCC5}27 & \cellcolor[HTML]{E7A295}49 & \cellcolor[HTML]{EDB9AF}37 & \cellcolor[HTML]{F9E7E3}13 \\  
 & CAR & \cellcolor[HTML]{E59F90}51 & \cellcolor[HTML]{E39585}56 & \cellcolor[HTML]{F4D6D0}22 & \cellcolor[HTML]{EFC1B8}33 & \cellcolor[HTML]{F2CCC5}27 & \cellcolor[HTML]{FFFEFD}1 \\  
 & CEN & \cellcolor[HTML]{E6A092}50 & \cellcolor[HTML]{E08C7B}61 & \cellcolor[HTML]{F4D4CD}23 & \cellcolor[HTML]{E9ACA0}44 & \cellcolor[HTML]{ECB7AD}38 & \cellcolor[HTML]{FAECEA}10 \\  
 & GHA & \cellcolor[HTML]{CC4125}117 & \cellcolor[HTML]{CC4125}102 & \cellcolor[HTML]{E8A89B}46 & \cellcolor[HTML]{E59D8E}52 & \cellcolor[HTML]{E39585}56 & \cellcolor[HTML]{E2EBFB}-16 \\  
 & MIL & \cellcolor[HTML]{E8A699}47 & \cellcolor[HTML]{E39585}56 & \cellcolor[HTML]{F2CEC7}26 & \cellcolor[HTML]{EAAEA2}43 & \cellcolor[HTML]{EEBFB5}34 & \cellcolor[HTML]{F8E5E1}14 \\  
 & MIS & \cellcolor[HTML]{DB7763}72 & \cellcolor[HTML]{D96F5A}76 & \cellcolor[HTML]{E9AA9D}45 & \cellcolor[HTML]{EAAEA2}43 & \cellcolor[HTML]{E9AA9D}45 & \cellcolor[HTML]{F7FAFE}-4 \\ %
 & NOR & \cellcolor[HTML]{F4D4CD}23 & \cellcolor[HTML]{EFC3BA}32 & \cellcolor[HTML]{F6DBD6}19 & \cellcolor[HTML]{EFC3BA}32 & \cellcolor[HTML]{F4D4CD}23 & \cellcolor[HTML]{F7DFDA}17 \\  
 & OUJ & \cellcolor[HTML]{DA735E}74 & \cellcolor[HTML]{D56048}84 & \cellcolor[HTML]{EBB4A8}40 & \cellcolor[HTML]{E6A092}50 & \cellcolor[HTML]{EAAEA2}43 & \cellcolor[HTML]{F7E1DD}16 \\  
 & PAL & \cellcolor[HTML]{EDBBB1}36 & \cellcolor[HTML]{E7A295}49 & \cellcolor[HTML]{F8E3DF}15 & \cellcolor[HTML]{EDBBB1}36 & \cellcolor[HTML]{F3D2CB}24 & \cellcolor[HTML]{FEFAF9}3 \\  
 & PAY & \cellcolor[HTML]{EEBFB5}34 & \cellcolor[HTML]{E29383}57 & \cellcolor[HTML]{F0C6BE}30 & \cellcolor[HTML]{E8A699}47 & \cellcolor[HTML]{ECB7AD}38 & \cellcolor[HTML]{F5D8D2}21 \\  
 & TAB & \cellcolor[HTML]{DE8472}65 & \cellcolor[HTML]{DF8876}63 & \cellcolor[HTML]{F2CEC7}26 & \cellcolor[HTML]{F3D0C9}25 & \cellcolor[HTML]{F0C5BC}31 & \cellcolor[HTML]{FEFAF9}3 \\  
 & TAT & \cellcolor[HTML]{D2553B}90 & \cellcolor[HTML]{D2563D}89 & \cellcolor[HTML]{F3D0C9}25 & \cellcolor[HTML]{EDBBB1}36 & \cellcolor[HTML]{EEBFB5}34 & \cellcolor[HTML]{D5E3F9}-23 \\  
 & TOR & \cellcolor[HTML]{F2CEC7}26 & \cellcolor[HTML]{EDB9AF}37 & \cellcolor[HTML]{F2CCC5}27 & \cellcolor[HTML]{EAB0A4}42 & \cellcolor[HTML]{EFC3BA}32 & \cellcolor[HTML]{F3D0C9}25 \\  
 \multirow{-14}{*}{Cloudy} & VIS & \cellcolor[HTML]{F7DFDA}17 & \cellcolor[HTML]{EDBBB1}36 & \cellcolor[HTML]{F4D4CD}23 & \cellcolor[HTML]{EFC3BA}32 & \cellcolor[HTML]{F5D8D2}21 & \cellcolor[HTML]{F5DAD4}20 \\ \bottomrule
\end{tabular}

\endgroup